\documentclass[review]{WileyASNA-v1}
\usepackage{amssymb,amsthm,amsmath,subcaption}
\usepackage{lipsum,natbib}

\usepackage[figuresright]{rotating}
\usepackage{siunitx}

\articletype{Article Type}%

\received{XX XXX XXXX}
\revised{XX XXX XXXX}
\accepted{XX XXX XXXX}

\begin{document}

\title{Tracing the Galactic Disk with $Gaia$ DR3: A Deep Study of Berkeley 17, 18, and 39 Open Star Clusters}

\author[1]{A. Ahmed}

\author[2]{W. H. Elsanhoury*}

\author[3]{D. C. Çınar}

\author[3]{S. Taşdemir}

\author[4]{R. Canbay}

\author[5,6]{A. A. Haroon}

\author[2]{M. S. Alenazi}

\authormark{A. Ahmed \textsc{et al}}

\address[1]{\orgdiv{Astronomy, Space Science and Meteorology Department  , Faculty of Science}, \orgname{Cairo University}, \orgaddress{\state{Giza, 12613, }, \country{Egypt}}}

\address[2]{\orgdiv{Department of Physics, College of Science}, \orgname{Northern Border University}, \orgaddress{\state{Arar}, \country{Saudi Arabia}}}

\address[3]{\orgdiv{Institute of Graduate Studies in Science, Programme of Astronomy and Space Sciences}, \orgname{Istanbul University}, \orgaddress{\state{Beyazıt, Istanbul 34116}, \country{Turkey}}}

\address[4]{\orgdiv{Faculty of Science, Department of Astronomy and Space Sciences}, \orgname{Istanbul University}, \orgaddress{\state{Beyazıt, Istanbul 34116}, \country{Turkey}}}

\address[5]{\orgdiv{Astronomy and Space Science Department, Faculty of Science}, \orgname{King Abdulaziz University}, \orgaddress{\state{Jeddah}, \country{Saudi Arabia}}}

\address[6]{\orgdiv{Astronomy Department}, \orgname{National Research Institute of Astronomy and Geophysics (NRIAG)}, \orgaddress{\state{Helwan 11421, Cairo}, \country{Egypt}}}

\corres{*W. H. Elsanhoury \\ \email{elsanhoury@nbu.edu.sa}}

\abstract{We report a detailed investigation of three intermediate-to-old age open clusters, Berkeley 17, Berkeley 18, and Berkeley 39 utilizing precise astrometric and photometric data from \textit{Gaia} DR3. Cluster membership was robustly determined through a probabilistic proper-motion analysis, yielding statistically significant samples of 600, 1042 and 907 stars, respectively. From the mean parallaxes of these members, we determine astrometric distances ranging from approximately 3.40 kpc for Berkeley 17 to 5.80 kpc for Be18. Isochrone fitting applied to the decontaminated color-magnitude diagrams constrains the cluster ages to 9.12 $\pm$ 1.00 Gyr, 3.36 $\pm$ 0.50 Gyr, and 5.10 $\pm$ 0.50 Gyr, respectively. Interstellar reddening spans a wide range, from $E(B-V) = 0.17$ mag in Berkeley 39 to 0.58 mag in Berkeley 17. Structural parameters derived from King model fits to the radial density profiles, combined with mass function analyses, indicate the clusters are dynamically relaxed systems with mass distributions broadly consistent with the canonical Salpeter slope. Our kinematic analysis reveals that Berkeley 17, 18 and 39 are part of the outer disk population.}



\maketitle

\section{Introduction}

\label{introduction}

Open clusters (OCs) are fundamental building blocks for understanding the formation and evolution of the Galactic disk. Their well-defined ages, distances, and chemical abundances allow them to serve as reliable tracers of the spatial and temporal evolution of the Milky Way (MW) \citep{Friel1995, Friel2013}. While young and intermediate-age clusters provide snapshots of ongoing and recent star formation, old OCs are particularly valuable because they are rare survivors of the early disk environment \textcolor{black}{\citep{Carraro1999a, Bragaglia2006a, LiuPang2019, Dias2021, Hunt2024}}. Their ages, often exceeding several Gyrs, make them crucial probes of the long-term dynamical and chemical evolution of the Galaxy. Old clusters preserve information about the conditions of the disk at the time of their formation and provide constraints on processes such as radial migration, disk heating, and chemical enrichment \citep{Salaris2004, Cant20b}. Studying these systems therefore offers unique insights into the history of the Galactic disk and bridges the gap between the properties of classical open clusters and globular clusters (GCs)\textcolor{black}{\citep{Perryma2026}}.

Star clusters are gravitationally bounded groups of stars that were born from the same parent interstellar cloud, and consequently, they share common chemical composition, space motion, and age. However, member stars can have different masses, and consequently can exist at different evolutionary stages depending on their masses \citep{Bostanci2015NGC6866,Bostanci2018}. The ages of stars in clusters can be easily determined from their observed \textcolor{black}{colour-magnitude diagrams (CMDs)}, unlike field stars, where it is difficult to get precise estimations of their ages \citep{Yontan2015NGC6811, Yontan2022, Cinar2024, TasdemirCinar2025, Tasdemir2025}. Also, investigating the mass distribution within star clusters is an important step that is needed to resolve many current debates, such as the predicted mass distributions first proposed by \cite{Salpeter_1955} in the clusters and the observed mass segregation in many stellar clusters. Due to the crucial importance of studying a large number of OCs, many catalogues have been published which include the measured physical parameters for thousands of galactic star clusters, e.g., \cite{Hunt2024, Cant20a, Cant20b, sampedro2017multimembership, kharchenko2013global,dias2014proper}.

The determination of the true members of the clusters and excluding field stars is an important step. The exclusion of true members or the addition of field stars may result in the underestimation or overestimation, respectively, of the derived physical properties of the clusters, e.g., their masses. As a result, many statistical algorithms were developed starting from the pioneer work of \citet{Vas58}, which use the observed stellar proper motions and spatial coordinates.

The current work presents a detailed photometric and dynamical study of three open star clusters, \textcolor{black}{Berkeley 17 (Be17), Berkeley 18 (Be18), and Berkeley 39 (Be39), using the most recent data of astrometric and photometric data from the $Gaia$ Release 3 database (DR3; \citet{GaiaDR3})}\footnote{\url{https://www.cosmos.esa.int/gaia}}.

\section{Target Clusters}

The three open clusters in this study, Berkeley 17, 18, and 39, were strategically selected to constitute a comprehensive sample of intermediate-to-old age systems. Such objects serve as fundamental tracers for the spatial and temporal evolution of the Galactic disk \citep{Friel1995, Friel2013}. Old open clusters are particularly valuable as they are rare survivors of the early disk environment and thus act as crucial probes of the long-term dynamical and chemical evolution of the Galaxy \citep{Carraro1999a, Bragaglia2006a}. Each cluster within our sample was chosen to contribute uniquely to this overarching scientific objective based on its distinct properties and place in the existing literature. Specifically, Be17 is renowned as one of the oldest known open clusters in the Milky Way; its advanced age, with estimates ranging from 8 to 10 Gyr, makes it an essential laboratory for constraining the formation timeline and history of the Galactic disk \citep{Bragaglia2006b, Salaris2004}. Be18, a rich cluster projected toward the Galactic anticenter, was chosen as an important probe of the intermediate-age population of the outer Galactic disk, a region critical for understanding disk formation models \citep{Kaluzny1997, Carraro1999b}. Be39 is another well-studied old open cluster whose combination of age and mass, along with its confirmed chemical homogeneity \citep{Bragaglia2012}, establishes it as a key calibrator for studies of disk chemical evolution. 

\begin{figure*}[ht]
\centering
\includegraphics[width=1\linewidth]{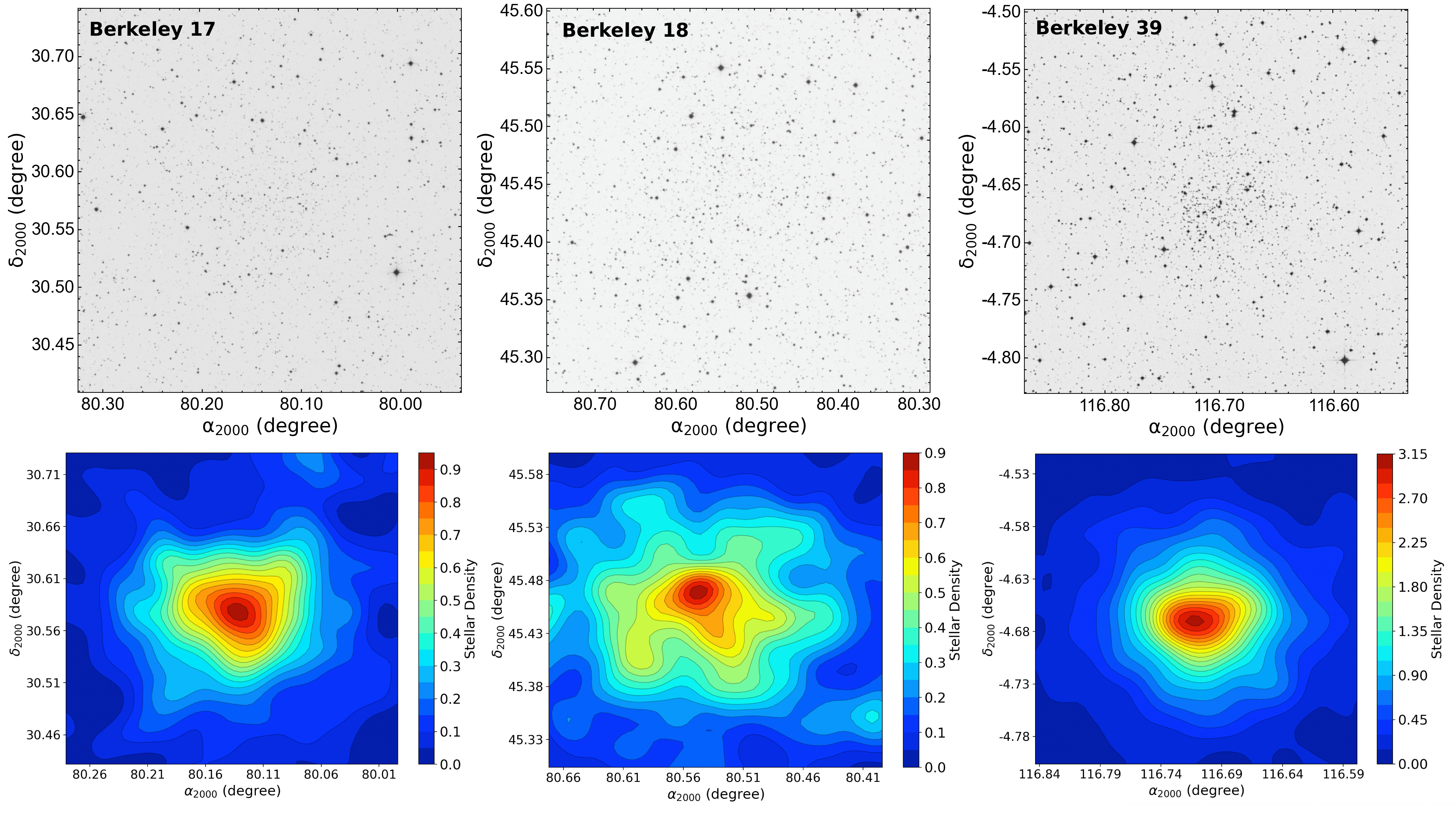}
\caption{DSS images illustrating the locations of the open clusters Be17, Be18,  and Be39 within the MW.}
\label{fig:finderchart}
\end{figure*}

\subsection{Berkeley 17}

Berkeley 17 is one of the oldest known open clusters in the MW, and consequently its study is crucial since it presents constrains on the formation time and the star formation history of the galactic disk. The earlier studies of \cite{Phelps1997, Phelps_etal1994} got estimation of its age of 12 Gyr which lies in the range of the youngest GCs. However, later studies such as \cite{Carraro1999a,  Carraro1999b, Salaris2004, Bragaglia2006b} got lower age estimates in the range of 8-10 Gyr, comparable to the thick-disk population. Photometric studies, both optical and near-infrared, have placed the cluster at a heliocentric distance of about 2.50-3.10 kpc with a relatively high reddening ($E(B-V)\approx$ 0.60 mag). Spectroscopic analyses indicate a subsolar metallicity of [Fe/H]=$-0.33\pm0.12$ dex \citep{Friel2002}. Its stellar population shows signatures of dynamical evolution, including mass segregation and tidal stripping, while $Gaia$ data have revealed a population of blue stragglers with a bimodal radial distribution, resembling those seen in GCs \citep{Bhattacharya2019}.

\subsection{Berkeley 18}

Berkeley 18 is a rich OC projected toward the Galactic anticenter, with an age of about 4-5 Gyr, slightly younger than M 67. It lies at a heliocentric distance of roughly 4.5-5.8 kpc and shows moderate extinction ($E(B-V) \geq 0.48$ mag). The cluster has a large angular extent ($\approx$26 arcmin), corresponding to a physical size of $\approx$44 pc, and is generally considered to have near-solar metallicity, though detailed spectroscopic studies remain scarce. Its properties make it an important probe of the intermediate-age population of the outer Galactic disk \citep{Kaluzny1997, Carraro1999b, Salaris2004}.

\subsection{Berkeley 39}

Berkeley 39 is another well-studied old OC, with age estimates ranging between 5 and 8 Gyr, a heliocentric distance of $\approx$4.2-4.5 kpc, and relatively low reddening value ($E(B-V) \approx$ 0.10 \textcolor{black}{mag}) \citep{Kaluzny1989,Tadross2003,Salaris2004, Krusberg2006, Vaidya2020}. High-resolution spectroscopic work has shown that, unlike massive globular clusters, Be39 does not display significant star-to-star variations in light elements such as Na and O, confirming that it is a chemically homogeneous, single-population system \citep{Bragaglia2012}. Generally, the cluster has a slightly subsolar metallicity ([Fe/H] $-0.26\pm0.09$ \textcolor{black}{dex}) \citep{Friel2002,Tadross2003,Paunzen2010,Bragaglia2012}. Its combination of age and mass makes it one of the key calibrators for studies of disk chemical evolution. Recent studies of \cite{Vaidya2020,Rao2021} revealed the presence of a population of blue straggler stars (BSS).

The current study presents a detailed photometric, spectroscopic, and dynamical study of three old and intermediate-age open star clusters, Be17, Be18, and Be39, using the most recent data of astrometric and photometric data from the $Gaia$ DR3 database \citep{bailer2023gaia, lindegren2021gaia, brown2021gaia,prusti2016gaia}. Figure \ref{fig:finderchart} presents the Digitized Sky Survey (DSS) images of the clusters, each covering a CCD field of view of approximately $20^{\prime} \times 20^{\prime}$. As the $50^{\prime} \times 50^{\prime}$ field of view is too large for these clusters and makes it difficult to distinguish the clusters from the surrounding field stars due to the field contamination, intensity plots were limited to a $20^{\prime}$ region. The $50^{\prime} \times 50^{\prime}$ data set was used for the detailed analyses.

\begin{table*}
\caption{Comparison of the estimated parameters in the literature for the investigated clusters:($\alpha$,~$\delta$ ; deg.), radius (r; arcmin), age (Myr), distance (d; pc), colour excess $E(B - V)$; mag), trigonometric parallax ($\varpi$; mas), proper motion components ($\mu_\alpha \cos \delta,~\mu_\delta$; mas $\rm yr^{-1}$), \textcolor{black}{the number of member stars (N), metallicity [Fe/H] and }the references.
\label{old_results}} 
\centering
\setlength\tabcolsep{3pt}
\scalebox{1.0}{
\begin{tabular}{cccccccccccc}
\hline
$\alpha$ & $\delta$ & $r$ & Age & $d$ &\it $E(B-V)$ & $ \varpi$ & $ \mu _\alpha \cos \delta$ & $ \mu _\delta$ & N & [Fe/H] & Ref.\\
(deg) & (deg) & (arcmin) & (Gyr) & (kpc) & (mag) & (mas) & (mas $\rm yr^{-1}$) &(mas $\rm yr^{-1})$& (stars) & (dex) &\\
\hline\hline
\multicolumn{11}{c}{Berkeley 17}\\
\hline
80.141 & 30.574 & 35.8 & 4.84 & 2.980 & --& 0.299 $\pm$ 0.121 & 2.537 $\pm$ 0.165 & $-$0.342 $\pm$ 0.132 & 586 & --&1\\
80.126 &30.574 & 4.20$^*$ & 9.98 & 3.135 & 0.555& 0.279 $\pm$ 0.121 & 2.611 $\pm$ 0.290 & $-$0.348 $\pm$ 0.231 & 254 & {$-0.069$} &2\\
80.130 & 30.574 & 4.26$^*$ & 7.24 & 3.341 & --& 0.298 $\pm$ 0.090 & 2.538 $\pm$ 0.128 & $-$0.346 $\pm$ 0.103 & 250 & --&3\\
80.130 & 30.574 & 4.26$^*$ & -- & 3.312 &--& 0.281 $\pm$ 0.106 & 2.618 $\pm$ 0.281 & $-$0.350 $\pm$ 0.176 & 261 &-- & 4\\
80.130 & 30.574 & 4.26$^*$ & 7.24 & 3.341 &--& 0.281 $\pm$ 0.106 & 2.618 $\pm$ 0.281 & $-$0.350 $\pm$ 0.176 & 254 &--&5\\
80.150 & 30.600 & 5 & 10.00 & 2.700 &0.58 &--& 4.73 $\pm$ 3.54 & $-$3.41 $\pm$ 3.39 & 127 &--&7\\
80.150 & 30.600 & 4.50 & -- & -- & -- &--& 3.60 $\pm$ 3.34 & $-$3.62 $\pm$ 2.27 & 105 &--&8\\
80.155 & 30.590 & 13.2 & 3.98 & 1.800 & 0.849 &--& 3.90 & $-$2.31 & -- &--&9 \\
\hline\hline
\multicolumn{11}{c}{Berkeley 18}\\
\hline
80.554 & 45.468 & 30.70 & 2.43 & 4.951 & --& 0.162 $\pm$ 0.091 & 0.782 $\pm$ 0.131 & $-$0.059 $\pm$ 0.112 & 926 & --&1\\
80.531 & 45.442 & 7.02$^*$ & 4.36 & 5.632 & --& 0.574 $\pm$ 0.036 & 0.771 $\pm$ 0.091 & $-$0.083 $\pm$ 0.065 & 250 &-- &3\\
80.531 & 45.442 & 7.02$^*$ & -- & 5.720 &--& 0.152 $\pm$ 0.088 & 0.849 $\pm$ 0.160 &$-$0.057 $\pm$ 0.140 & 317 & --&4\\
80.531 & 45.442 & 7.02$^*$ & 4.36 & 5.632 & -- &0.152 $\pm$ 0.088 & 0.849 $\pm$ 0.160 &$-$0.057 $\pm$ 0.140 & 253 & --&5\\
80.503&45.442 & 16.38 & 0.12 & -- & --& 0.221 $\pm$ 0.017 & 0.756 $\pm$ 0.342 & $-$0.104 $\pm$ 0.221 & 78 &-- &6\\
80.550 & 45.400 & 10.88 & 4.27 & 5.800 & 0.46 &--& 0.650 $\pm$ 4.040 &$-$4.460 $\pm$ 4.500 & 470 & -- &7\\
80.550 & 45.400 & 7.0 & -- & -- & -- & -- & 0.330 $\pm$ 0.100 & $-$4.520 $\pm$ 0.900 & 234 & -- &8\\
80.602 & 45.405 & 16.5 & 4.27 & 5.800 & 0.45 &--& $-$3.58 & $-$5.50 & -- & -- &9\\
\hline\hline
\multicolumn{11}{c}{Berkeley 39}\\
\hline
116.707 & $-$4.672 & 39.40 & 4.55 & 3.840 & --& 0.231 $\pm$ 0.072 & $-$1.732 $\pm$ 0.087 & $-$1.631 $\pm$ 0.075 & 1203 & --&1\\
116.696& $-$4.658 & 3.42$^*$ & 6.47 & 4.356 & 0.127& 0.207 $\pm$ 0.117 & $-$1.724 $\pm$ 0.217 & $-$1.648 $\pm$ 0.150 & 526 & {$-0.109$}&2\\
116.702 & $-$4.665 & 3.30$^*$ & 5.62 & 3.968 & --& 0.226 $\pm$ 0.067 & $-$1.728 $\pm$ 0.073 & $-$1.632 $\pm$ 0.069 & 525 &-- &3\\
116.702 & $-$4.665 & 3.30$^*$ & -- & 4.461 &--& 0.201 $\pm$ 0.103 &$-$1.730 $\pm$ 0.179 & $-$1.645 $\pm$ 0.129 & 543 & --& 4\\
116.702 & $-$4.665 & 3.30$^*$ & 5.62 & 3.968 &--& 0.201 $\pm$ 0.103 & $-$1.730 $\pm$ 0.179 & $-$1.645 $\pm$ 0.129 & 530 & --&5\\
116.695&$-$4.655 & 12.78 & 12.30 & -- & --& 0.268 $\pm$ 0.045	 & -1.679$\pm$0.216	&$-$1.643$\pm$0.213 & 234 & -- &6\\
116.700 & $-$4.668 & 4.22 & 7.94 & 4.780 & 0.12 & -- & 0.52 $\pm$ 7.93 & $-$5.09 $\pm$ 7.51 & 105 & --&7\\
116.675 & $-$4.600 & 4.50 & -- & -- & -- &--& $-$2.97 $\pm$ 1.08 & $-$2.57 $\pm$ 3.73 & 92 & --&8\\
116.707 & $-$4.680 & 9.00 & 3.16 & 4.836 & 0.042 &--& $-$2.50 & $-$1.01 & -- & --&9\\

\hline\\
\end{tabular}}
\newline
References: (1)\cite{Hunt2024}; (2)\cite{Dias2021}; (3)\cite{Pog21};(4)\cite{Cant20a};(5)\cite{Cant20b}; (6)\cite{LiuPang2019});(7) \cite{sampedro2017multimembership}; (8)\cite{dias2014proper}; (9)\cite{kharchenko2013global}. 
\\ $^*$ These values are $r_{50}$ or the radii containing half the members. 
\end{table*}


\section{Data}

The current work made use of the published astrometric and photometric data in the third release of the $Gaia$ project \cite{GaiaDR3}. $Gaia$ project provides the five-parameter astrometry for around 1.8 billion sources, together with their location on the sky ($\alpha$, $\delta$), trigonometric parallaxes (Plx; mas) ($\varpi$), \textcolor{black}{and the proper motion components ($\mu_\alpha \cos \delta$ and $\mu_\delta$).} The limiting magnitude of the proper motion (PM; mas $\rm yr^{-1}$) components ($\mu_\alpha \cos \delta$, $\mu_\delta$) is \textit{G} = 21 mag. Up to 0.02-0.03 mas $\rm yr^{-1}$ (at \textit{G} $<15$ mag), 0.07 mas $\rm yr^{-1}$ (at \textit{G} $\approx$ 17 mag), 0.50 mas $\rm yr^{-1}$ (at \textit{G} $\approx~20$ mag), and 1.40 mas $\rm yr^{-1}$ (at \textit{G} = 21) are the uncertainties in the corresponding PM components. $\approx$ 0.02–0.03 mas for sources with \textit{G} $<15$ mag, $\approx$ 0.07 mas for sources with $G=17$ mag, $\approx~0.50$ mas at $G=20$ mag, and $\approx~1.30$ mas at $G=21$ mag are the uncertainties in the trigonometric parallax values \citep{bailer2023gaia}. $Gaia$ DR3 is showing significant improvement over $Gaia$ DR2 \textcolor{black}{\citep{Evans_2018}}; trigonometric parallax measurement accuracy has increased by $30\%$, and PM measurement accuracy has increased by a factor of 2.00. Clusters’ data was extracted by a $50'$ radius areas around the clusters' central coordinates, so that it exceeded the reported radii. Within these regions, astrometric and photometric data from $Gaia$ DR3 catalog for Be17, Be18, and Be39, were obtained, comprising 89647, 95903, and 64154 stars, respectively. The photometric and astrometric errors in the $Gaia$ DR3 catalogue exhibit a strong dependence on the $G$ magnitude. For the brightest stars ($G<15$ mag), the typical uncertainties are of the order of $\sim 0.002-0.003$ mag in the $G$ band, $\sim 0.02-0.03$ mas in trigonometric parallaxes, and $\sim 0.02-0.03$ mas~yr$^{-1}$ in proper motions. At intermediate brightness ($G\sim20$ mag), the uncertainties rise to approximately 0.01 mag, 0.5 mas, and 0.5 mas~yr$^{-1}$, while at the faintest limit ($G\simeq 21$ mag), they can reach up to 0.02 mag in $G$, 1.3 mas in parallaxes, and 1.4 mas~yr$^{-1}$ in proper motions \citep{brown2021gaia}. These error characteristics were fully incorporated into the membership selection and subsequent isochrone-fitting analyses. The final cleaned datasets, therefore, contain high-quality photometric and astrometric measurements for all stars brighter than the completeness limit in each cluster. A summary of the mean photometric uncertainties per magnitude interval is provided in Table~\ref{tab:photometric_errors}.

\subsection{Photometric Completeness Analysis}

\begin{figure}[ht]
\centering
\includegraphics[width=0.95\linewidth]{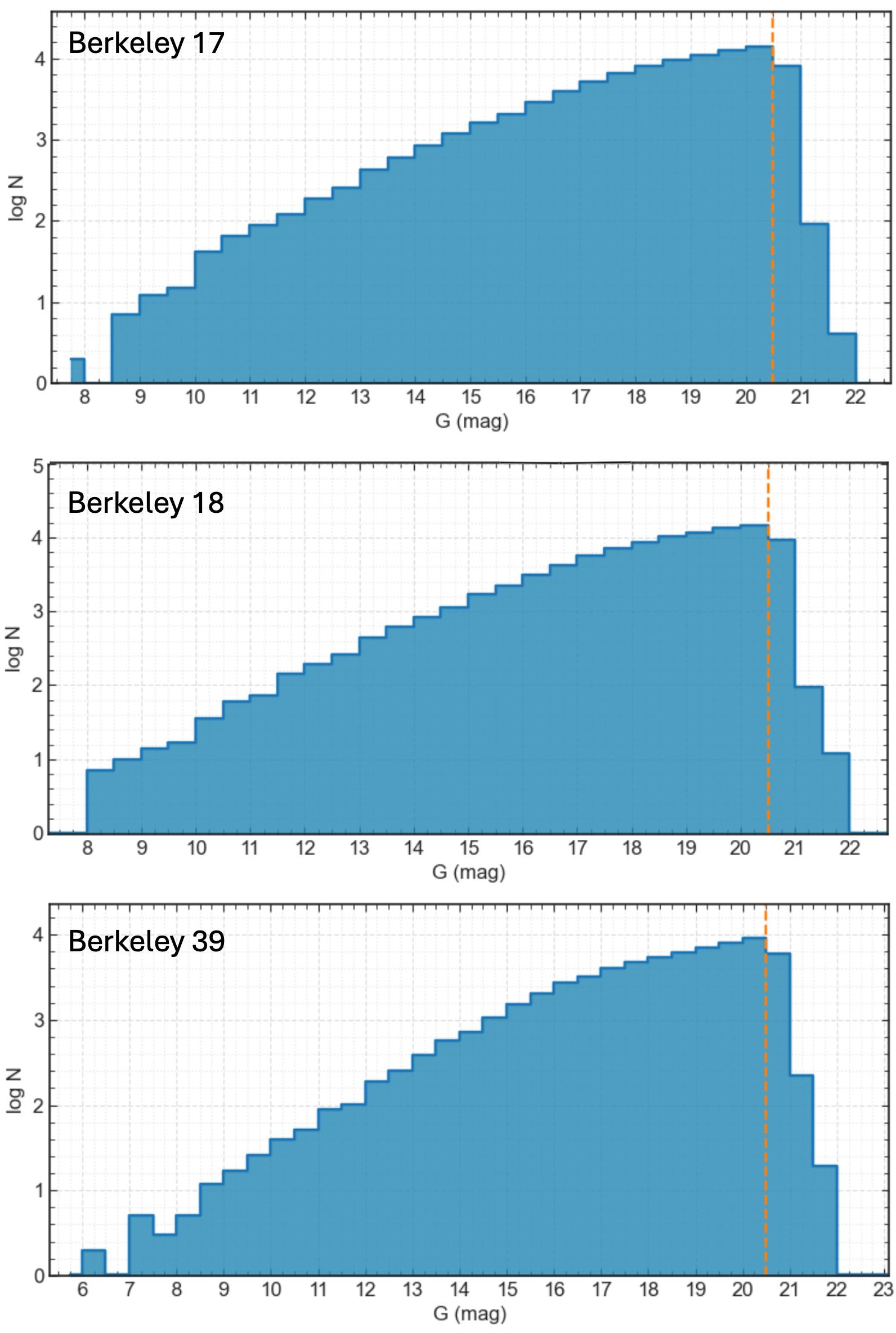}
\caption{Completeness histograms for the open clusters Be17, Be18, and Be39. The red dashed line in each panel indicates the estimated photometric completeness limit in \textcolor{black}{the $Gaia~G$-apparent magnitude}.}
\label{fig:completness}
\end{figure}

A reliable determination of the photometric completeness threshold is essential for accurate derivation of the fundamental parameters of open clusters. In order to assess these limits for the Berkeley sample, we generated $G$-band magnitude distributions for the stars located within the cluster regions. The star counts show a monotonic increase toward fainter magnitudes until a characteristic peak is reached, after which the numbers decline because of incompleteness. The magnitude corresponding to this turnover was adopted as the photometric limit and obtained as $G=20.50$ \textcolor{black}{mag} for each cluster (see Figure \ref{fig:completness}). Stars fainter than these thresholds were excluded from further analysis to avoid systematic biases. After applying the completeness corrections, the remaining stellar samples consist of 74290, 78896, and 53105 stars for Be17, Be18, and Be39, respectively.

\begin{table*}[h]
\centering
\caption{Mean photometric uncertainties in $Gaia$ apparent magnitudes ($G$) and colour \textcolor{black}{indices} ($G_{\rm BP}-G_{\rm RP}$) for stars in the direction of the three OCs. The values are presented as a function of $G$ magnitude intervals.}
\label{tab:photometric_errors}
\begin{tabular}{c|ccc|ccc|ccc}
\hline
 & \multicolumn{3}{c|}{\textbf{Berkeley 17}} & \multicolumn{3}{c|}{\textbf{Berkeley 18}} & \multicolumn{3}{c|}{\textbf{Berkeley 39}}  \\
\cmidrule(lr){2-4} \cmidrule(lr){5-7} \cmidrule(lr){8-10} $G$ (mag) & $N$ & $\sigma_{\rm G}$ & $\sigma_{G_{\rm BP}-G_{\rm RP}}$ & $N$ & $\sigma_{\rm G}$ & $\sigma_{G_{\rm BP}-G_{\rm RP}}$ & $N$ & $\sigma_{\rm G}$ & $\sigma_{G_{\rm BP}-G_{\rm RP}}$  \\
\hline \hline
6--14   & 1472 & 0.0029 & 0.0055 & 1569 & 0.0029 & 0.0061 & 1443 & 0.0028 & 0.0054  \\
14--15  & 1728 & 0.0029 & 0.0055 & 1674 & 0.0028 & 0.0055 & 1502 & 0.0029 & 0.0054  \\
15--16  & 3229 & 0.0029 & 0.0067 & 3304 & 0.0029 & 0.0067 & 3068 & 0.0029 & 0.0064  \\
16--17  & 5828 & 0.0030 & 0.0107 & 6266 & 0.0030 & 0.0099 & 5376 & 0.0030 & 0.0096  \\
17--18  & 10369 & 0.0032 & 0.0210 & 11322 & 0.0031 & 0.0194 & 8026 & 0.0031 & 0.0186  \\
18--19  & 16151 & 0.0037 & 0.0483 & 17302 & 0.0035 & 0.0440 & 10884 & 0.0035 & 0.0419  \\
19--20  & 22087 & 0.0055 & 0.1106 & 23527 & 0.0048 & 0.0927 & 14255 & 0.0049 & 0.0913  \\
20--21  & 27189 & 0.0117 & 0.2491 & 28721 & 0.0105 & 0.2255 & 18048 & 0.0113 & 0.2236  \\
21--23  & 1594 & 0.0272 & 0.4339 & 2218 & 0.0253 & 0.4287 & 1552 & 0.0275 & 0.3869  \\
    \hline
Total/Error & 89647   &0.0032&	\textcolor{black}{0.0210} & 95903   & 0.0031 &	0.0194 & 64154    &0.0031 &	0.0186   \\
   
\hline
\end{tabular}
\end{table*}



\section{Structure analysis}

\subsection{Clusters' centers}

Redetermination of OCs centers can be performed by constructing histograms of the right ascension ($\alpha$) and declination ($\delta$) of all stars in the data file obtained from the $Gaia$ DR3 dataset, which is crucial for enhancing our comprehension of cluster formation and dynamics.
Histograms give an approachable and straightforward method that is particularly helpful in preliminary analyses and instructional contexts, even while more complex methods such as Kernel Density Estimation (KDE) or machine learning models offer more precision. Peaks of the Gaussian fits mark the region with the highest star density that will be considered as the cluster's center.

\begin{figure*}[ht]
    \centering
    \includegraphics[width=0.99\linewidth]{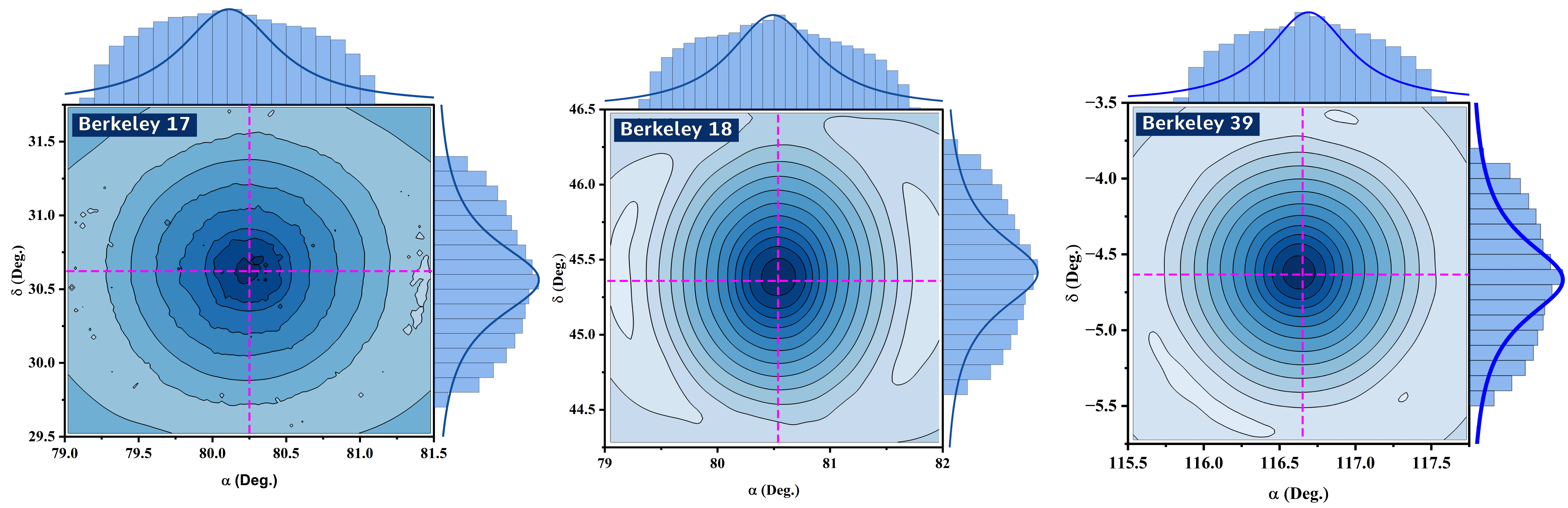}
    \caption{The two-dimensional Gaussian modeling was employed to determine the central coordinates of the stellar density enhancements in $\alpha$ and $\delta$. The resulting density maps and contour distributions reveal the spatial structure and stellar concentration patterns of the Berkeley OCs investigated in this study. \textcolor{black}{Pink dashed lines indicate the center of the clusters.}}
    \label{fig:centers}
\end{figure*}

To achieve this, we created $\alpha$ and $\delta$ histograms , as seen in Figure \ref{fig:centers}, separating the extracted region into a number of bins with identical sizes, and used Gaussian fitting. Therefore, the cluster centers exist at ($\alpha$, $\delta$) $\&$ ($l$, $b$) are listed here with Table \ref{all_results}.

%
%

\subsection{Radial Density Profiles}

\begin{table}
\caption{The structural parameters of the OCs under study derived from the \citet{King1962} fitting. Quoted uncertainties correspond to the $1\sigma$ confidence intervals obtained from the posterior distributions.}
 \renewcommand{\arraystretch}{1.3}
\label{King_para}
\scalebox{0.9}{
\begin{tabular}{lccc} 
\hline
\textbf{Parameters} & \textbf{Berkeley 17} & \textbf{Berkeley 18} & \textbf{Berkeley 39}  \\
\hline\hline
$ r_{\rm cl}$ (arcmin) 
& $12.61^{+0.49}_{-0.49}$ 
& $19.96^{+0.48}_{-0.48}$ 
& $12.96^{+0.47}_{-0.47}$ \\

$ r_{\rm c}$ (arcmin)
& $3.56^{+0.47}_{-0.40}$ 
& $7.04^{+0.58}_{-0.52}$ 
& $3.31^{+0.30}_{-0.27}$ \\ 

$\rho_0$ $\rm (stars\,arcmin^{-2})$ 
& $17.63^{+2.12}_{-1.92}$ 
& $16.63^{+1.13}_{-1.08}$ 
& $28.35^{+2.59}_{-2.39}$ \\

$\rho_{\rm bg}$ $\rm (stars\,arcmin^{-2})$ 
& $9.25^{+0.04}_{-0.04}$ 
& $9.76^{+0.04}_{-0.04}$ 
& $6.60^{+0.03}_{-0.03}$ \\ 

$C$ 
& \textcolor{black}{3.54}
& \textcolor{black}{2.84 }
& \textcolor{black}{3.92} \\
$C_{Literature}$ 
& \textcolor{black}{4.91$\rm ^a$}
& \textcolor{black}{3.49$\rm ^a$}
& \textcolor{black}{3.88$\rm ^a$} \\
$\delta_c$ 
& 2.91 
& 2.70 
& 5.29 \\
\hline
\multicolumn{4}{l}{\footnotesize References: (a) \citet{Hunt2024}.}
\end{tabular}}
\end{table}

\begin{figure*}
\centering
\includegraphics[width=1\linewidth]{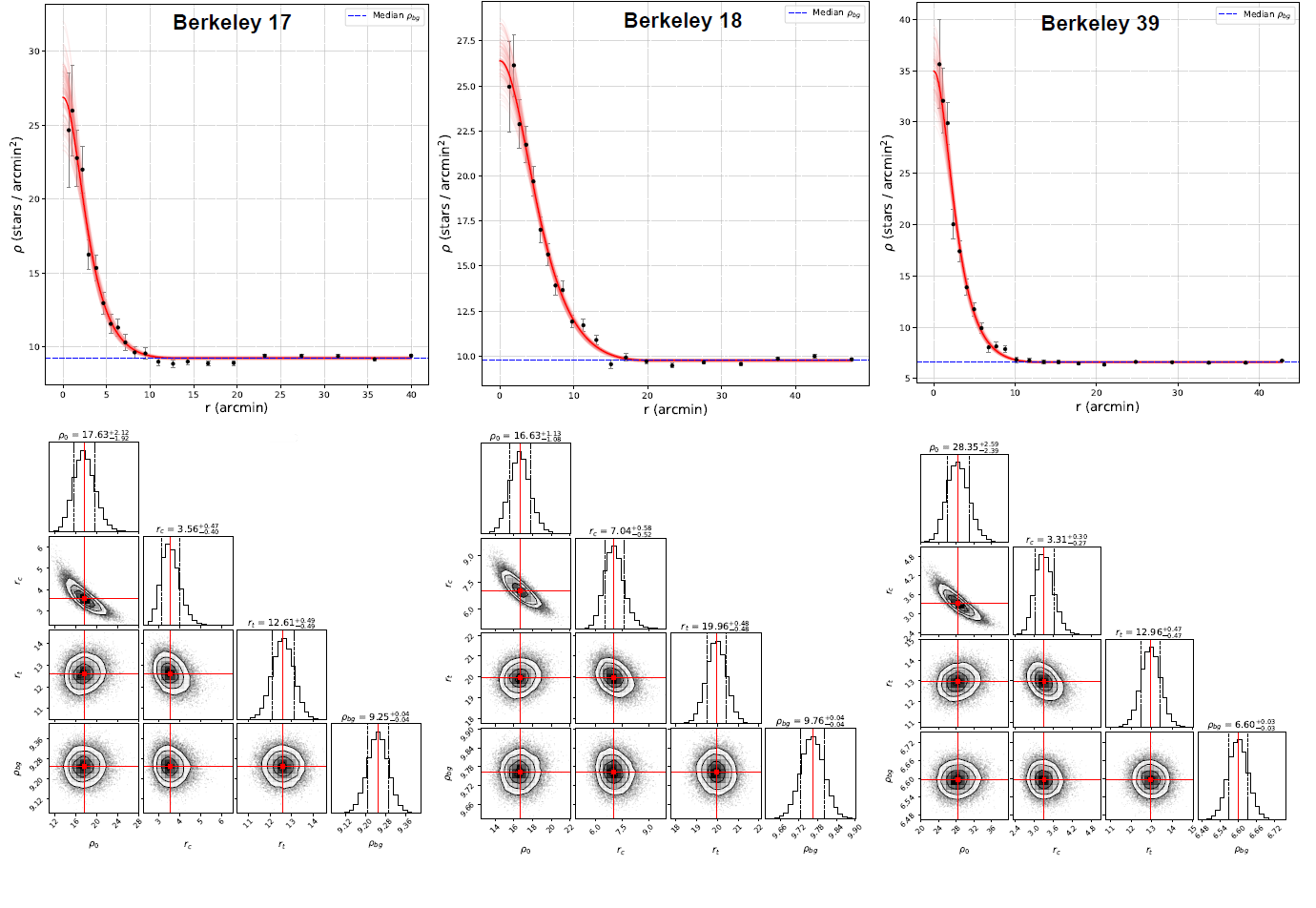}
\caption{The RDPs of the investigated OCs represented by the blue solid line. The dashed red line represents the \cite{King1962} fitting while the red solid line and blue dashed line the background density $\rho_{\rm bg}$ and the uncertainties of the background density, respectively. 
\label{rdps}}
\end{figure*}

Details of the internal structure of the clusters can be inferred from their radial density profiles (RDPs), which provide insight into the degree of central concentration and the spatial extent of the stellar systems.

The RDPs were constructed using the selected cluster member stars, as illustrated in Figure~\ref{rdps}.
The observed field was divided into concentric annuli centred on the cluster centre, with a fixed radial step of 1 arcmin. The stellar surface density in each annulus was calculated as:
\begin{equation}
\rho_i=\frac{N_i}{A_i},
\end{equation}
where $N_i$ is the number of stars within the $i$-th annulus and $A_i$ is its area, expressed in arcmin\textsuperscript{2}.
The associated uncertainties were estimated assuming Poisson statistics, such that $\sigma_{\rho_i} = \sqrt{N_i}/A_i$, which naturally accounts for sampling fluctuations, particularly in the inner regions of the clusters.

The cluster radius was defined as the distance at which the stellar surface density becomes indistinguishable from the surrounding field star density \citep{tadross2005analytical}.
To derive the structural parameters, the observed RDPs were fitted with the empirical King surface-density model \citep{King1962}, expressed as:
\begin{equation}
\rho_r=\rho_{\rm bg}+\frac{\rho_0}{1+(r/r_c)^2},
\end{equation}
where $\rho_{\rm bg}$ denotes the background stellar density, $\rho_0$ the central density above the background level, and $r_c$ the core radius, defined as the distance at which the surface density drops to half of its central value.



The parameters were estimated via maximum likelihood, using the log-likelihood expression:
\begin{equation}
\ln \mathcal{L} = -\sum_i 
\left( \frac{\rho_i - \rho_{i,\mathrm{Model}}}{\sigma_{\rho_i}} \right)^2 .
\end{equation}
Sampling was performed with the \texttt{emcee} MCMC algorithm \citep{Foreman-Mackey2013}, adopting uniform priors. Convergence was assessed through the Gelman--Rubin diagnostic \citep{gelman1992}. Figure \ref{rdps} shows the resulting RDPs of all clusters together with the corresponding model fits.

Further insight into the cluster structure was obtained by computing the density contrast parameter ($\delta_c$) and the concentration parameter ($C$).
The density contrast parameter is defined as:
\begin{equation}
\delta_c = 1 + \frac{\rho_0}{\rho_{\rm bg}},
\end{equation}
and provides a quantitative measure of the degree to which the cluster stands out against the background stellar field.
Higher values of $\delta_c$ indicate a more prominent and centrally enhanced stellar system, consistent with dynamically evolved open clusters.
The concentration parameter, defined as $C = r_{\rm cl}/r_c$ \citep{King1966}, describes the compactness of the core region relative to the overall cluster size.

\textcolor{black}{We found $C$ = 3.54, 2.84, and 3.92 for Be17, Be18, and Be39, }respectively, indicating moderate to high degrees of central concentration among the studied clusters. The derived cluster radii, best-fitting King model parameters, density contrast values, and concentration parameters are summarised in Table~\ref{King_para}. For our clusters, we obtain $\delta_c$ = 2.91, 2.70, and 5.29 for Be17, Be18, and Be39, respectively, indicating that Be39 stands out most prominently against the field, consistent with its strongly centrally concentrated structure.\textcolor{black}{ To place our structural parameters in context, we compare our concentration parameter with corresponding literature values reported for the same clusters. \citet{Hunt2024} list $C_{\rm Literature}=4.91$, 3.49, and 3.88 for Be17, Be18, and Be39, respectively (see Table~\ref{King_para}). Our concentrations preserve the same relative ranking and differ only moderately from these determinations, which is plausibly attributable to methodological differences (e.g., the adopted definition of the outer cluster radius $r_{\rm cl}$,background estimation, radial binning, and the membership selection used to construct the RDP).}


\section{The Photometric Analysis}

\subsection{Membership Determination}\label{gaia-membership}

The probable members of the investigated clusters were identified following the statistical approach of \cite{Yad13} that is based on the procedure of \cite{Bal98}, which made use of the observed proper motion. Proper motion provides a better description of the stellar motion over two dimensions compared with the radial velocity, and it is also less affected by the orbital motion in unrecognized binaries \citep{Tian98}. 
In this approach, we plot the vector point diagram (VPD) using the $\alpha$ and $\delta$ components of the proper motion ($\mu_\alpha \cos\delta$, $\mu_\delta$) over different intervals of the $\it G$ magnitudes, and then the cluster zone is identified as the region of the highest star density in the brightest $\it G$ interval, $\it G\leq$ 17 \textcolor{black}{mag} in the present study. Next, stars that lie in this region over all $\it G$ \textcolor{black}{magnitudes} are labelled as cluster stars, while those that lie outside this region are labelled as field stars.

The calculation of the membership probabilities of the stars that exist within the estimated radius of each cluster starts by computing the frequency distribution for the cluster’s stars ($\phi_c^\nu$ ) and for the field stars ($\phi_f^\nu$). The cluster’s frequency distribution is computed by:
\begin{equation}
\phi_{c}^{\nu}= \frac{1}{2\pi\sqrt{\left(\sigma_{c}^{2}+ \epsilon_{xi}^{2}\right)\left(\sigma_{c}^{2}+\epsilon_{yi}^{2}\right)}} \times e^{-\frac{1}{2}\left[\frac{\left(\mu_{xi}-\mu_{xc}\right)^{2}}{\sigma_{c}^{2}+\epsilon_{xi}^{2}}+\frac{\left(\mu_{yi}-\mu_{yc}\right)^{2}}{\sigma_{c}^{2}+\epsilon _{yi}^{2}}\right]}\,, 
\end{equation}
where $\sigma_c$ is the intrinsic proper motion dispersion of cluster member stars that was given a value of 0.075 mas $\rm yr^{-1}$ \citep{Sin20}. This value is based on the assumption of \cite{Gir89} that radial velocity dispersions within open clusters are equal to 1.00 km $\rm s^{-1}$ \citep{Sin20}. In the equation, $\mu_{xc}$ and $\mu_{yc}$ are the cluster's PM centre, $\mu_{xi}$ and $\mu_{yi}$ are the $\alpha$ and $\delta$. components of the $\rm i^{th}$ star, and $\epsilon_{xi}$ and $\epsilon_{yi}$ are the corresponding observed errors.

Next, the frequency distribution of the field stars was computed using:
\begin{align}
\phi_{f}^{\nu}=\frac {1}{2\pi\sqrt{\left(1-\gamma^{2}\right)}\sqrt{\left(\sigma_{xf}^{2}+\epsilon_{xi}^{2}\right)\left(\sigma_{yf}^{2}+\epsilon_{yi}^{2}\right)}} \times \nonumber\\
&\hspace{-5.0cm}e^{-\frac{1}{2\left(1-\gamma^{2}\right)}\left[\frac{\left(\mu_{xi}-\mu_{xf}\right)^{2}}{\sigma_{xf}^{2}+\epsilon _{xi}^{2}} - \frac{2\gamma\left(\mu_{xi}-\mu_{xf}\right)\left(\mu_{yi}-\mu_{yf}\right)}{\sqrt{\left(\sigma_{xf}^{2}+\epsilon _{xi}^{2}\right)\left(\sigma_{yf}^{2}+\epsilon _{yi}^{2}\right)}} + \frac{\left(\mu_{yi}-\mu_{yf}\right)^{2}}{\sigma_{yf}^{2}+\epsilon _{yi}^{2}}\right]}\, 
\end{align}
where $\mu_{xf}$ and $\mu_{yf}$ are the average value of $\alpha$ and $\delta$. components of the proper motions of all field stars, while their corresponding dispersions are $\sigma_{xf}$ and $\sigma_{yf}$, and $\gamma$ is the correlation coefficient computed by:
\begin{equation} \gamma = \frac{\left(\mu_{xi}-\mu_{xf}\right)\left(\mu_{yi}-\mu_{yf}\right)}{\sigma_{xf}\sigma_{yf}} \,, \end{equation}

The distribution of all stars can be derived from:
\begin{equation}
\Phi = \left(n_{c}\cdot\phi_{c}^{\nu}\right) + \left(n_{f}\cdot\phi_{f}^{\nu}\right)\,, 
\end{equation}
where $n_c$ and $n_f$ denote the normalized number of stars in the cluster and field regions, such that $n_c + n_f= 1$. Accordingly, the probability of $\rm i^{th}$ star’s membership is given by:
\begin{equation}
P_{\mu}(i) = \frac{\phi_{c}(i)}{\phi(i)}\,. \end{equation}

The parameters used in the membership test for the investigated clusters and the radii of the selected cluster zones ($r_{\rm zone}$) are listed in Table \ref{member-parameters}. We got 600, 1042, and 907 members for Be17, Be18, and Be39, respectively, with probabilities equal to or greater than 50 \citep{Bisht2021, Bisht2022b, Bisht2025}.\textcolor{black}{The 50\% membership threshold was adopted following the reliability criterion of \cite{Yad13}, ensuring a balance between completeness and purity in the member sample.}

The effectiveness of the membership is computed using the formula of \cite{shao1996effectivity}:
\begin{equation}
E=1-\frac{N \sum_{i=1}^{N} [P(i)(1-P(i))]}{\sum_{i=1}^{N} P(i) \sum_{i=1}^{N} (1-P(i))}\, , \end{equation}
where higher values of $E$ represent high effectiveness of the procedure. The values of $E$ in literature vary between 0.2 to 0.9, and it has an optimum value at 0.55 \citep{shao1996effectivity}. The values of the effectiveness of membership determination for the three clusters are also listed in Table \ref{member-parameters}, which reflect the high effectiveness of member selection in the present work.

\begin{center}
\begin{table}[htbp]
\centering
\caption{Parameters used in the membership selection.}
\label{member-parameters}
\scalebox{0.95}{
\begin{tabular}{lrrr} 
\hline
\textbf{Parameters} & \textbf{Berkeley 17} & \textbf{Berkeley 18} & \textbf{Berkeley 39}  \\

\hline\hline

$\mu_{xc}$ $\rm (mas.yr^{-1}$) & 2.48 & 0.72 & $-$1.73  \\

$\mu_{yc}$ $\rm (mas.yr^{-1}$) & $-$0.37 & $-$0.10 & $-$1.67  \\

$\mu_{xf}$ $\rm (mas.yr^{-1}$) & 1.28 & 0.93 & $-$1.61  \\ 

$\sigma_{xf}$ $\rm (mas.yr^{-1}$) & 3.00 & 2.89 & 3.21  \\

$\mu_{yf}$ $\rm (mas.yr^{-1}$) & $-$2.24 & $-$2.00 & $-$0.72  \\

$\sigma_{yf}$ $\rm (mas.yr^{-1}$) & 3.84 & 3.97 & 4.65  \\

$n_f$ & 0.91 & 0.95 & 0.84  \\

$n_c$ & 0.09 & 0.05 & 0.16  \\

$r_{zone}$ $\rm (mas.yr^{-1}$) & 0.30 & 0.20 & 0.2  \\

$E$ & 0.69 & 0.59 & 0.86  \\

\hline 

\end{tabular}}
\end{table}
\end{center}


\subsection{Astrometric Properties of the OCs}

\begin{figure}
    \centering
\includegraphics[width=0.8\linewidth]{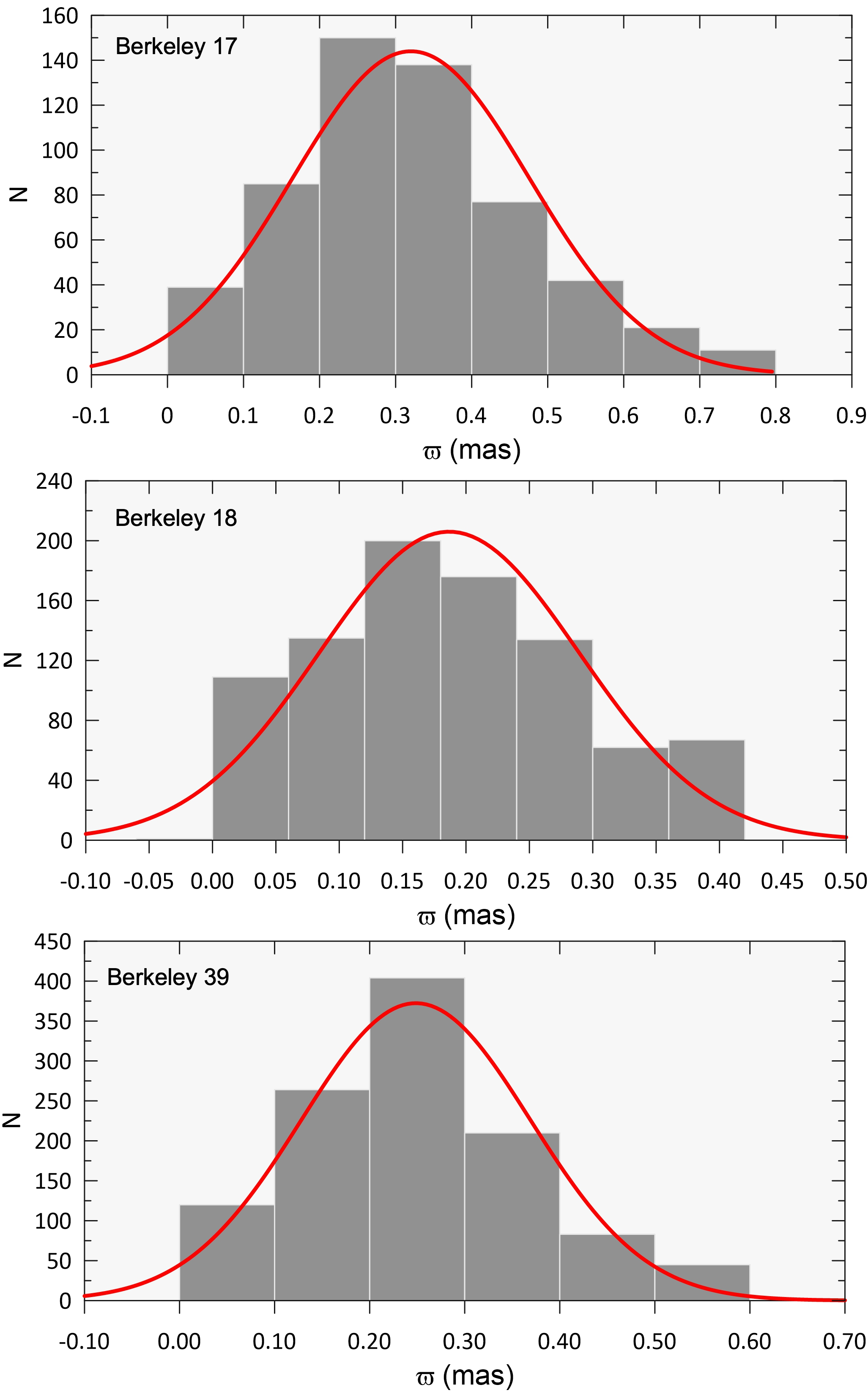}
    \caption{Normalized trigonometric parallax distribution of stars with membership probability $P \geq 0.5$ for Be17, Be18, and Be39. The red curve represents the best-fit Gaussian to the distribution.}
    \label{fig: Plx}
\end{figure}

\begin{figure}[ht]
    \centering
    \includegraphics[width=0.99\linewidth]{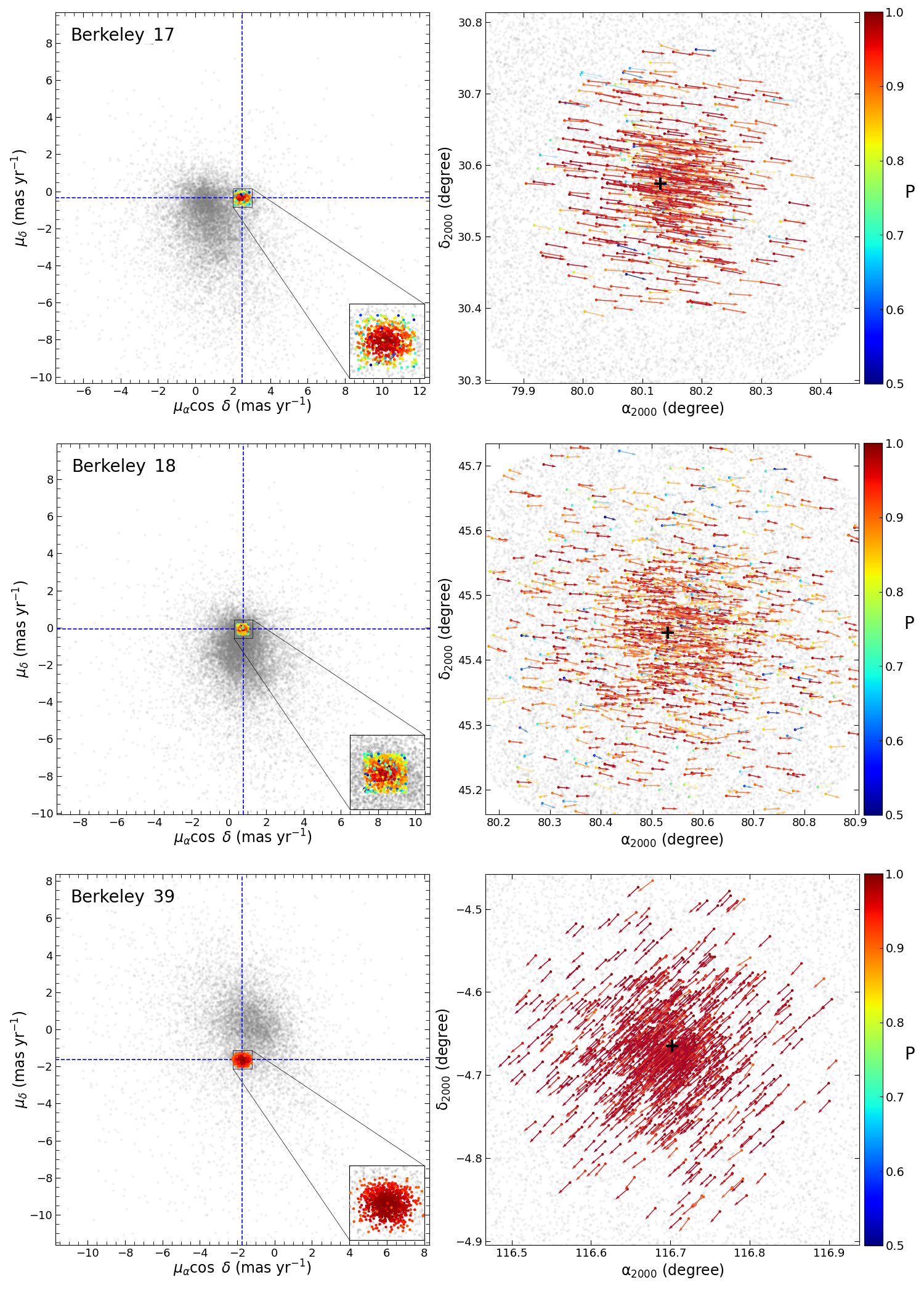}
    \caption{VPDs of the three OCs showing stellar positions with proper-motion vectors in equatorial coordinates. Cluster centers are marked by black plus signs, and blue dashed lines denote median proper-motion components.}
    \label{fig: PM}
\end{figure}

\textcolor{black}{In order to get estimates of the average proper-motion components in both directions, and the average parallax for every cluster, we constructed histograms for the proper-motion components and parallaxes of all member stars, and then we searched for the best-fit Gaussian profiles of the constructed histograms. The mean values and standard deviations of the best fit profiles were assigned to the cluster as shown in Figure \ref{fig: Plx}.
We found that the mean proper-motion components in both directions of Be17 (2.528 $\pm$ 0.192, $-$0.337 $\pm$ 0.142; mas yr$^{-1}$), Be18 (0.768 $\pm$ 0.153, $-$0.066 $\pm$ 0.137; mas yr$^{-1}$), and Be39 ($-$1.726 $\pm$ 0.108, $-$1.631 $\pm$ 0.092; mas yr$^{-1}$), where the created and the constructed histograms are devoted with Figure \ref{fig: PM}.}

Also, the mean trigonometric parallaxes of probable members are equal to 0.297 $\pm$ 0.146 (Be17), 0.173 $\pm$ 0.116 (Be18), and 0.237 $\pm$ 0.097 (Be39) in mas, as shown in Figure \ref{fig: Plx}, and their corresponding distances are 3373 $\pm$ 68, 5767 $\pm$ 119, and 4214 $\pm$ 57, pc with respective manner of Be17, Be18, and Be39 OCs. Our results agree fairly well with the distances reported in the literature for these clusters, as shown in Table \ref{old_results}.

\subsection{CMDs, Ages and Distances}\label{CMD-Age-Dist}

In OC studies, color-magnitude diagrams (CMDs) are very useful since they show the red giant, MS turn-off point, and other traits that correlate to star age and composition. The CMDs for thousands of OCs have been enhanced with previously unheard-of astrometric precision with the release of $Gaia$ Data Release 3 \textcolor{black}{\citep{GaiaDR3}}, enabling smoother diagrams and improved cluster membership determination.

Comparing the observed CMDs, constructed using the observed photometric data of the selected members of each cluster with the theoretically computed isochrone, enables us to derive many important physical characteristics of the clusters, like age, distance, reddening, and metallicity. In the present work, we made use of the theoretical isochrones of \cite{marigo2017new} computed assuming a solar metallicity, $Z=0.0152$ for the OCs Be18, and theoretical isochrones computed assuming $Z=0.008$ for OCs Be17 and Be39.

In the present study, the metallicity of each cluster was taken into account when selecting the appropriate theoretical isochrones. The conversion from spectroscopically determined [Fe/H] values to the heavy-element mass fraction ($Z$) was carried out following the prescription implemented in the \texttt{isodist} package\footnote{\url{https://github.com/jobovy/isodist/blob/master/isodist/Isochrone.py}}, which has been widely adopted in recent works \citep{Elsanhoury2025, Haroon2025, Alzahrani2025b, Alzahrani2025a} and is fully compatible with the PARSEC evolutionary models \citep{Bressan12}. In this scheme, an auxiliary parameter $Z_{\rm x}$ is obtained from
\begin{equation}
\label{eq:Zx}
Z_{\rm x} = 10^{\rm [Fe/H] + \log \left( Z_\odot / (1 - 0.248 - 2.78 \times Z_\odot) \right)},
\end{equation}
\textcolor{black}{\citep{2023AJ....165...79Y,2023AJ....166..263G}} and subsequently transformed into the global heavy-element abundance:
\begin{equation}
\label{eq:Z}
Z = \frac{Z_{\rm x} - 0.2485 \, Z_{\rm x}}{1 + 2.78 \, Z_{\rm x}}.
\end{equation}
Here, the reference solar metallicity is adopted as $Z_\odot = 0.0152$. Based on this formalism, the heavy-element contents of Be17, Be18, and Be39 were evaluated from their respective iron abundances of [Fe/H]~$=$~-0.29, 0.00, and -0.29 dex, which were then used to select the corresponding PARSEC isochrones for age determination.

\begin{figure*}[ht]
    \centering
\includegraphics[width=0.95\linewidth]{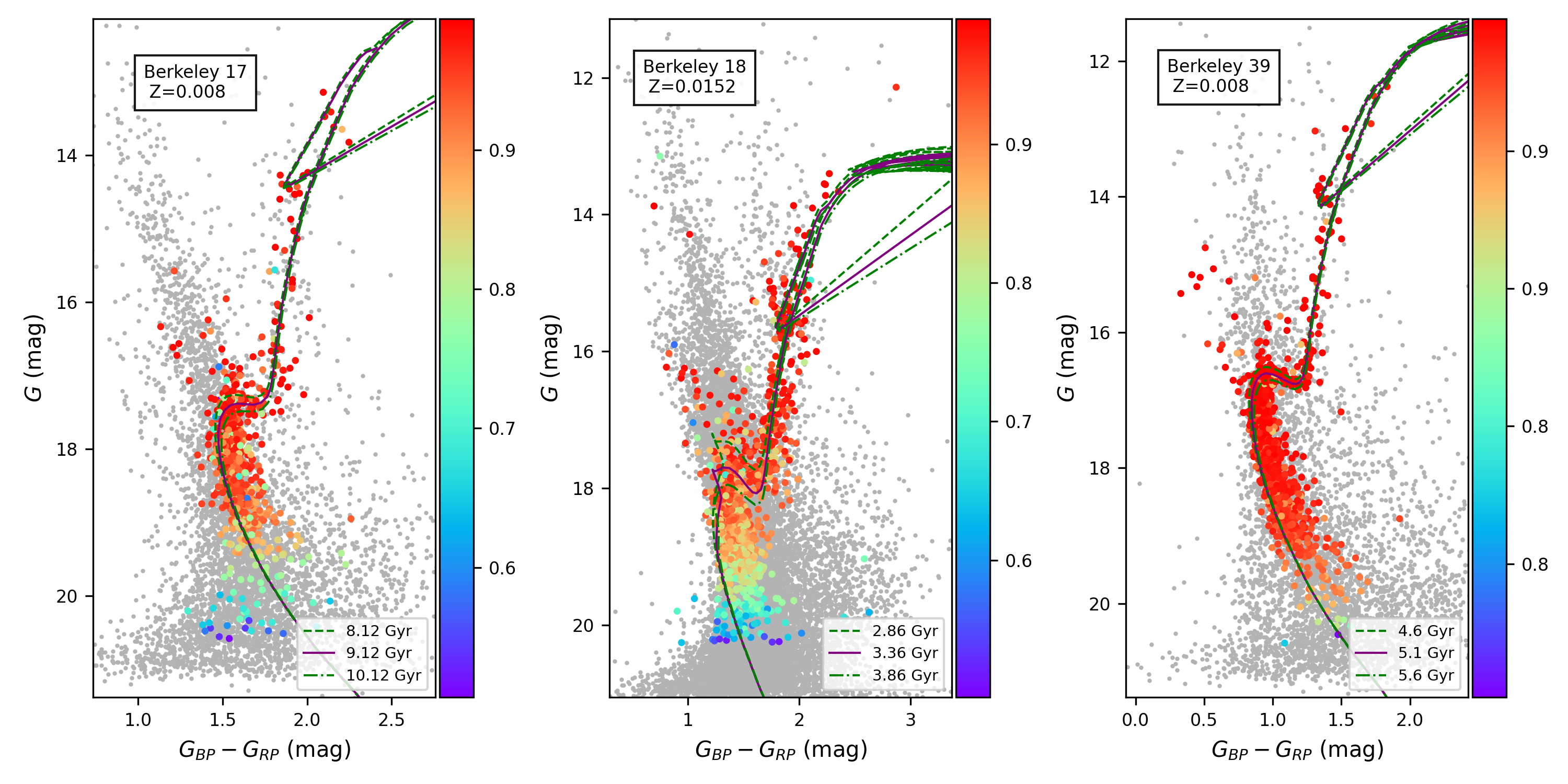}
    \caption{\textcolor{black}{CMDs} of the open clusters Be17, Be18, and Be39. Probable cluster members with $0.5 \leq P \leq 1$ are shown in distinct color codes, while field stars are represented by grey circles. The best-fitting {\sc PARSEC} isochrones are plotted in purple, with their corresponding uncertainty ranges indicated in green.}
    \label{Berkeleys-CMDs}
\end{figure*}

Figure \ref{Berkeleys-CMDs} shows the {\it G } versus $(G_{\rm BP} - G_{\rm RP})$ CMDs for the three studied clusters and the best fit isochrones that obtained by visual inspection, i.e. comparing the observed CMD of each cluster with isochrones of different ages, where each isochrone was shifted vertically and horizontally until the best fit isochrone was obtained. The age of the best-fit isochrones were assigned to the clusters.
The isochrone provide absolute magnitudes, $M_{\rm G}$, and intrinsic colours, $(G_{\rm BP} - G_{\rm RP})$\citep{Canbay2025b}, computed for different stellar masses at the same age, and consequently, the difference between the observed stellar magnitudes and the absolute magnitudes of the best fit ischrone represent the observed distance modulus of the cluster, $m_G-M_G$, and the difference between the observed stellar colours and the corresponding values of the isochrone represent the colour excess of the cluster, $E(G_{\rm BP} - G_{\rm RP})$ \citep{Ak2016NGC6819, Ak2015CV}.

As shown in Fig. \ref{Berkeleys-CMDs}, the best fit isochrones for Be17, Be18, and Be39 are 9.12 $\pm$ 1.00, 3.36 $\pm$ 0.50, and 5.10 $\pm$ 0.50 Gyr, respectively. We found that, for the OCs Be17 and Be39, changing the adopted metallicity value has no effect on the estimated age, but it caused a slight change of the estimated values of the distance moduli and reddening in agreement with \cite{Carraro1999a}. For Be17, reducing the adopted metallicity led to an increase in the distance modulus and colour excess by 0.20, 0.15 mag, respectively, and consequently caused a reduction of the distance by $\approx$160 pc, which is less than the estimated uncertainty.
The estimated distance moduli and colour excesses of Be17 (13.90, 0.75; mag), of Be18 (15.20, 0.60; mag), and of Be39 (13.73, 0.22; mag). The corresponding $E(B-V)$ colour excesses are obtained using $E(B-V)=0.775$ $\times E(G_{\rm BP}-G_{\rm RP})$ \citep{cardelli1989relationship} which are equal to 0.47 mag (Be17), 0.47 mag (Be18), and 0.17 mag (Be39). 
Also, the line-of-sight extinction coefficients are estimated by $A_G=2.74 \times E(B-V)$ \citep{casagrande2018use,zhong2019substructure,Canbay_2023}, which are equal to 1.59 mag (Be17), 1.27 mag (Be18), and 0.47 mag (Be39). 
The photometric distances of the clusters can be obtained using their estimated distance moduli by the following relation
\begin{equation}
d=10^{((m-M)_{\rm obs}-A_{\rm G}+5)/5}\,.
\end{equation}

The photometric distances of Be17, Be18, and Be39 \textcolor{black}{are 2894 $\pm$ 271, 6098 $\pm$ 715, 4493 $\pm$ 391 pc}, respectively. The measured photometric distances agree within uncertainties with the corresponding astrometric distances for Be18 and Be39. 
Our estimated ages of Be17 and Be18 agree within uncertainties with the previously reported values of \cite{Pog21, Cant20b,sampedro2017multimembership}, but it is higher than the reported value of \cite{Hunt2024}. 

The estimated age of Be39 agrees within uncertainties with the previously reported value of \cite{Hunt2024,Pog21, Cant20b}, but it's lower than that of \cite{sampedro2017multimembership} and higher than that of \cite{kharchenko2013global}.

Also, our estimated distances mostly agree within uncertainties with the previously reported distances, see Table \ref{old_results}. Using the measured parallax-based distance, the distances of the clusters from the Galactic plane, $Z_\odot$, and their projected distances from the Sun, $X_\odot$ and $Y_\odot$, and their distances from the Galactic centre $R_{\rm GC}$ were computed by the formula given by \citet{Tuncel19}:
\begin{align}
X_{\odot} = d \cos b \cos l\,, Y_{\odot} = d \cos b \sin l \,, Z_{\odot} = d \sin b \, 
\end{align}

\begin{equation}
R_{\rm GC} = \sqrt {R_{\odot}^{2}+(d \cos b)^{2}-2R_{\odot}d \cos b \cos l}\,, 
\end{equation}
where $d$ is the distance to the Sun in pc, $R_\odot $ is the distance between the Sun and the Galactic centre. $R_\odot $ was assigned a value of 8.20 $\pm$ 0.10 kpc as reported by \cite{bland2019galah}. The computed values of $R_{\rm GC}$ are \textcolor{black}{10.88, 13.95 and 11.61  kpc for Be17}, Be18, and Be39, respectively.

\begin{table*}[ht]
\centering
\caption{Summary of the results obtained in the present work.
\label{all_results}}
\scalebox{0.85}{
\begin{tabular}{lccc} 
\hline
Parameter & \textbf{Berkeley 17} & \textbf{Berkeley 18} & \textbf{Berkeley 39}  \\
\hline
\hline
\textcolor{black}{$\alpha $ (deg)} & $\rm 5^h 20^m 29.06^s \pm  0.39^s$  & $\rm 5^h 21^m 59.66^s \pm 0.46^s$ & $\rm 7^h 46^m 45.96^s \pm  0.39^s$ \\ 
\textcolor{black}{$\delta $ (deg)} & $\rm 30^{\circ} 34' 2.65'' \pm 4.95''$ &     $\rm 45^{\circ} 25' 7.47'' \pm 4.73''$   & $\rm -4^{\circ} 39' 59.25'' \pm 5.86''$  \\   
$\ell \;\&\;b$ (deg) & 176.6589 \& $-$3.6870 & 164.5956 \& 4.9977 & 223.5330 \& 10.0771 \\ 
$r$ (arcmin) & 12.66 & 19.95 & 12.63  \\ 

\hline
\multicolumn{4}{c}{Astrometric Parameters} \\
\hline   
$\varpi$ (mas) & 0.297 $\pm$ 0.146 & 0.173 $\pm$ 0.116 & 0.237 $\pm$ 0.097  \\ 
\textcolor{black}{$d_\varpi$ (pc)}& 3373 $\pm$ 68 & 5767 $\pm$ 119 & 4214 $\pm$ 57  \\  
\textcolor{black}{$\mu _{ \alpha }\cos \delta \; (mas\;yr^{-1})$} & 2.528 $\pm$ 0.192 & 0.768 $\pm$ 0.153 & $-$1.726 $\pm$ 0.108  \\ 
\textcolor{black}{$\mu_\delta \; (mas \;yr^{-1})$} & $-$0.337 $\pm$ 0.142 & $-$0.066 $\pm$ 0.137 & $-$1.631 $\pm$ 0.092  \\
$N_{stars}$ & 600 & 1042 & 907  \\
\hline
\multicolumn{4}{c}{Astrophysical Parameters} \\
\hline
$(m-M)_{obs}$ (mag) & 13.90 $\pm$  0.10& 15.20 $\pm$ 0.10 & 13.73 $\pm$ 0.10 \\
$E( G_{\rm BP}-G_{\rm RP})$ (mag) & 0.75 $\pm$ 0.03& 0.60 $\pm$ 0.05 & 0.22 $\pm$  0.02\\
$E(B-V)$ (mag) & 0.581& 0.465 & 0.170 \\ 
$ A_{G}$ (mag) & 1.593 & 1.274 & 0.467 \\
\textcolor{black}{d($m-M$) (pc)} & 2894 $\pm$ 271 & 6098 $\pm$ 715 & 4493 $\pm$ 391 \\
$Z$     &  0.008  &  0.0152    &    0.008     \\
$$[Fe/H]$$ (dex)   &  -0.29    &    0.00        &    -0.29     \\
Age (Gyr) & 9.12  $\pm$ 1.00& 3.36 $\pm$ 0.50& 5.10  $\pm$ 0.50\\ 
\hline
\multicolumn{4}{c}{Distances from Galactic Plane and Galactic Centre} \\
\hline
\textcolor{black}{$X_{\odot}$ (kpc) }& -2.88 $\pm$ 0.05 & -5.85 $\pm$ 0.08 & -3.21 $\pm$ 0.06  \\ 
\textcolor{black}{$Y_{\odot}$ (kpc)} & 0.16 $\pm$ 0.02 & 1.61 $\pm$ 0.04 & -3.05 $\pm$ 0.06  \\
\textcolor{black}{$Z_{\odot}$ }(kpc) & -0.18 $\pm$ 0.02 & 0.53 $\pm$ 0.02 & 0.78 $\pm$ 0.03  \\
\textcolor{black}{$R_{GC}$(kpc)} & 10.58 $\pm$ 0.10 & 13.95 $\pm$ 0.13 & 11.61 $\pm$ 0.07  \\
\hline 
\end{tabular}}
\end{table*}

\section{Luminosity and Mass Functions}
\label{sec_LF_and_MF}

The luminosity function (LF) describes the distribution of stars in a cluster according to their \textcolor{black}{absolute magnitudes}, whereas the mass function (MF) represents their distribution in terms of stellar masses \citep{Tasdemir2023, YontanCanbay2023}.

\begin{figure*}
\centering
\includegraphics[width=0.8\linewidth]{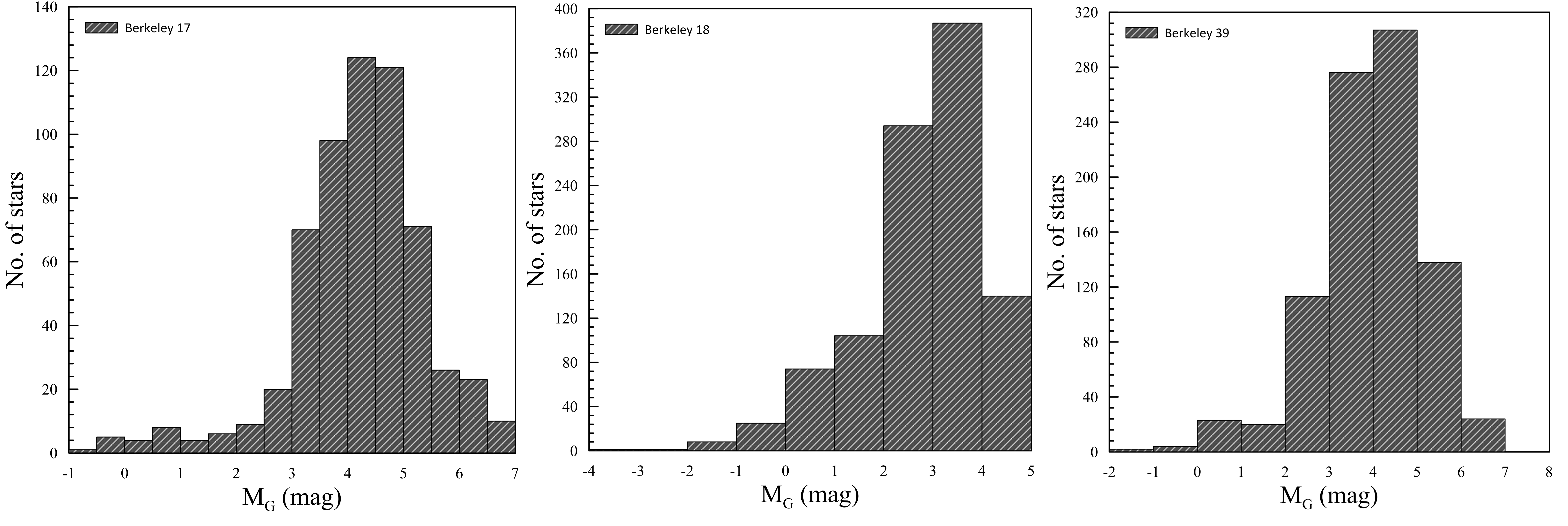}
\caption{Luminosity \textcolor{black}{functions} of Be17, Be18, and Be39.}
\label{figure:LF}
\end{figure*}

Since all members of an OC originate simultaneously from the same molecular cloud under similar physical conditions, these systems serve as excellent laboratories for studying stellar distributions with respect to magnitudes and masses. The LF of the Be17, Be18, and Be39 members are shown in Figure~\ref{figure:LF}. The LF displays a slight upward trend, suggesting that the cluster has retained its low-and intermediate-mass stars, an uncommon property among older open clusters in the MW. 

A clear link exists between the LF, MF, and the mass–luminosity relation (MLR; \cite{eker2024}). In this study, we adopted the relation between absolute magnitude $M_{\rm G}$ and stellar mass $M/M_\odot$ from \citet{Evans_2018}, corresponding to a metallicities and ages sorted in Table \ref{all_results}. This calibration covers the magnitude ranges $-0.755 \leq M_{\rm G}~{\rm (mag)} \leq 6.679$, $-3.062 \leq M_{\rm G}~{\rm (mag)} \leq 5.058$, and $-1.351 \leq M_{\rm G}~{\rm (mag)} \leq 6.850$ with respective manners of Be17, Be18, and Be39 OCs. By applying a polynomial of the fourth order with our retrieved MLR, therefore the coefficients are listed in Table \ref{MF_results} and is represented by: 
\begin{equation}
M_{\rm c} = a_0 + a_1 \times M_{\rm G} + a_2 \times M_{\rm G}^2 + a_3 \times M_{\rm G}^3 + a_4 \times M_{\rm G}^4,
\label{Eq: ML}
\end{equation}
where $a_0,~a_1,~a_2,~a_3$, and $a_4$ are polynomial coefficients that we get from the fitting.

\begin{table}[ht]
\centering
\caption{The MFs and their derived parameters.}
\label{MF_results}
\scalebox{0.80}{
\begin{tabular}{lccc} 
\hline
Parameter & \textbf{Berkeley 17} & \textbf{Berkeley 18} & \textbf{Berkeley 39}  \\
\hline
\hline
\textcolor{black}{$N_{\rm evolved}$} & 123 & 315 & 115 \\
\textcolor{black}{$\langle M/M_\odot \rangle$$_{\rm evolved}$} & 0.974 & 1.445 & 1.148\\
\textcolor{black}{$M_{C}/M_\odot(\rm evolved)$} & 120 & 455 & 132  \\
$N_{MS}$ & 477 & 727 & 792 \\
\textcolor{black}{$\langle M/M_\odot \rangle$$_{MS}$} & 0.872 & 1.219 & 0.976\\ 
\textcolor{black}{$M_{C}/M_\odot (MS)$} & 416 & 887 & 773  \\ 
$M_{C}$ ($M_\odot$) & 536 & 1342 & 905 \\
$\alpha$ & 2.44 $\pm$ 0.06 & 2.72 $\pm$ 0.06 & 2.25 $\pm$ 0.07  \\ 
$a_o$ & 0.885 $\pm$ 0.002 & 1.425 $\pm$ 0.002 & 1.102 $\pm$ 0.001  \\
$a_1$ & 0.082 $\pm$ 0.001 & -0.095 $\pm$ 0.001 & 0.090 $\pm$ 0.001  \\
$a_2$ & -0.002 $\pm$ 0.001 & 0.126 $\pm$ 0.001 & -0.018 $\pm$ 0.001  \\
$a_3$ & -0.006 $\pm$ 0.001 & -0.051 $\pm$ 0.001 & -0.005 $\pm$ 0.001  \\
$a_4$ & 0.001 $\pm$ 0.00 & 0.005 $\pm$ 0.001 & 0.001 $\pm$ 0.001  \\
\hline
\end{tabular}}
\end{table}

\begin{figure}[b]
\centering
\includegraphics[width=0.99\linewidth]{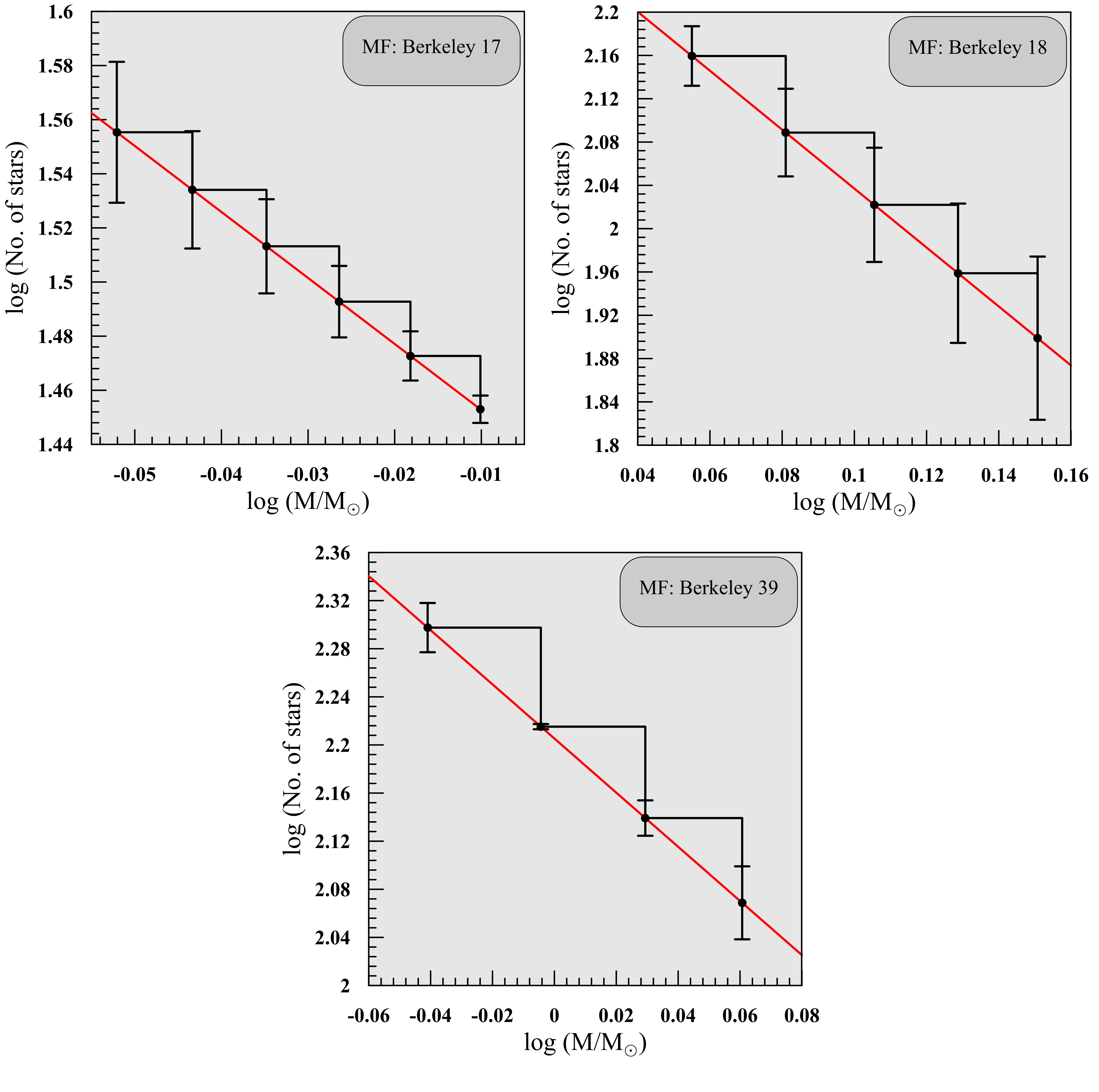}
\caption{Mass function of the three cluster members, fitted with the power-law relation of \citet{Salpeter_1955} (red line), which was used to determine the slope ($\alpha$) for the cluster.}
\label{figure:MLR}
\end{figure}

Once this relation is established, the total stellar mass of the cluster within a given magnitude range can be estimated by substituting the observed $M_{\rm G}$ values into the polynomial expression. This approach enables mass determination in cases where direct measurements are impractical, making use of the accurately calibrated MLR derived from isochrone models. The corresponding values are summarized in Table \ref{MF_results}.

OCs contain stars spanning a wide range of masses, from very low-mass stars to massive ones. This diversity makes them key environments for probing the initial mass function (IMF), which describes the primordial stellar mass distribution. The IMF has been extensively studied in previous works \citep{Phelps_janes_1993, Piskunov_2004}.The observed total mass of each cluster could be estimated by integrating the stellar masses over the observed luminosity range. The MLRs were obtained using the selected isochrone date in the brightness \textcolor{black}{ranges -1 $\leq M_G (mag)\leq$ 7, -3 $\leq M_G(mag) \leq$ 5, and -1.5 $\leq M_G(mag) \leq$ 7 for Be17}, Be18, and Be39, respectively. Table \ref{MF_results} shows the average, total mass, and the coefficients of the MLR for each investigated cluster.




\citet{Salpeter_1955} first introduced the IMF as a measure of the number of stars ($dN$) distributed over a logarithmic mass interval ($dM$) for a given stellar mass $M$. It is expressed as $dN/dM = M^{-\alpha}$, where $\alpha$. For Salpeter’s law, $\alpha$ is equal to 2.35. Here, $\alpha$ represents the slope of the MF. This slope, particularly for stars with $M > 1~M_{\odot}$ for our main-sequence stars (MS), serves as an indicator of dynamical evolution. Salpeter’s power law highlights the steep decline in stellar numbers with increasing mass. The MF analysis was only limited to the MS, the locations of the turn of main sequence point on the CMDs were found at $G$ magnitude values equal to 17.39, 17.70 and 16.60 for Be17, Be18 and Be39, respectively, where the corresponding value $M_G$ values are 3.49, 2.50 and 2.87. Consequently, MS have magnitudes higher than or equal to these values, while the post MS have magnitudes lower than these values. Also, due to the observed incompleteness of the $M_G$ distributions shown in Fig. \ref{figure:LF}, stars with $M_G$ \textcolor{black}{values fainter 4.5, 4 and 5 mag} for Be17, Be18 and Be39, respectively, were not included in the MF analysis too. This was done to avoid due to the signifiant underestimation of star numbers in the bins with fainter magnitudes which may cause significant effect on the derived values of the slopes of the MFs. Consequently, the MF analyses were \textcolor{black}{limited to 3.49 $\leq M_G (mag)\leq$ 4.5, 2.5 $\leq M_G(mag)\leq$ 4, 2.87 $\leq M_G(mag)\leq$ 5 for} Be17, Be18 and Be39, respectively.

For this study, MF calculations were performed for MS stars within the mass ranges 0.878 $\leq M(M_{\odot})_{Be~17} \leq$ 0.986, 1.100 $\leq M(M_{\odot})_{Be~18} \leq$ 1.450, and 0.870 $\leq M(M_{\odot})_{Be~39} \leq$ 1.19,  which includes a representative stellar sample. By applying a least-squares fit with MF data as seen in Figure~\ref{figure:MLR} as a straight line, we derived a slopes are $\alpha_{Be~17} = 2.44 \pm 0.06 $, $\alpha_{Be~18} = 2.72 \pm 0.06$, and $\alpha_{Be~39} = 2.25 \pm 0.07$, consistent with Salpeter’s value within the associated uncertainties. Additionally, the coefficients of the least-squares as well as their total and average masses are listed in Table \ref{MF_results}.

\section{Evolution times and the kinematical structures}
\subsection{Dynamical Relaxation Time}

The relaxation time ($T_{\rm relax}$) corresponds to the characteristic timescale over which stars in a cluster experience sufficiently weak gravitational encounters with their neighbors to alter their velocity vectors in a cumulative manner. Over this interval, the cluster gradually evolves from an initially non-equilibrium state toward a dynamically relaxed configuration. Following the formulation of \citet{Spitzer1971}, $T_{\rm relax}$ can be estimated as:
\begin{equation}
\label{Eq:t_relax}
T_{\rm relax} = \frac{8.9 \times 10^5 N^{1/2} R_{\rm h}^{3/2}}{\langle M_C \rangle^{1/2} \log(0.4N)},
\end{equation}
where $N$ is the total number of members, $R_{\rm h}$ is the half-mass radius in parsecs and $\langle M_C \rangle$ denotes the mean stellar mass of the system. To derive the half-mass radius, we adopted the empirical relation of \citet{Sableviciute_2006}:
\begin{equation}
R_{\rm h} = 0.547~\times~r_{\rm c}~\times \left( \frac{r_{\rm cl}}{r_{\rm c}} \right)^{0.486},
\end{equation}

The numericals values of $R_h$ (pc) as  3.46 $\pm$ 0.34, 7.51 $\pm$ 0.47 and 4.07$\pm$ 0.29, for Be17, Be18, Be39 OCs, respectively, and $r_{\rm cl}$ (pc) being the cluster radius determined from the RDP analysis (see Table~\ref{King_para}). Based on this approach, the relaxation times are estimated as $T_{\rm relax}$ (Myr) = 62\textcolor{black}{$\pm$2}, 199\textcolor{black}{$\pm$7}, and 86\textcolor{black}{$\pm$2} for Be17, Be18, Be39 OCs, respectively. 

The ratio $\tau = \text{age} / T_{\rm relax}$ serves as a useful diagnostic of the dynamical stage of a cluster systems with $\tau \gg 1$ are considered dynamically relaxed. Therefore, our program OCs serve as relaxed ones. The evaporation timescale, defined as $\tau_{\rm ev} \simeq 10^2 T_{\rm relax}$ (Myr) \citep{2001ApJ...553..744A}, characterizes the time required for the cluster to lose all of its members through internal stellar encounters. In this process, preferential loss of low-mass stars occurs through the Lagrangian points, typically at low ejection velocities \citep{koposov2008}. Table~\ref{tab:full_parameters} represent the corresponding values of $R_{\rm h}$, $T_{\rm relax}$, $\tau$, and $\tau_{ev}$.

\subsection{CP}
On the celestial sphere, the PM vectors of stars in a co-moving group, such as an open cluster, appear to converge due to perspective at a position called the Convergent position (CP) \citep{Elsanhoury2018, Bisht2022a, Elsanhoury2025, Elsanhoury2025b}. If a cluster of stars is travelling nearly parallel across space, the CP is the point at which the observed PMs of the cluster appear to converge when projected onto the sky. Mathematics-wise, this concept is related to vector motion equations and spherical trigonometry \citep{Galli_2012}.

\begin{figure}[ht]
\centering
\includegraphics[width=0.99\linewidth]{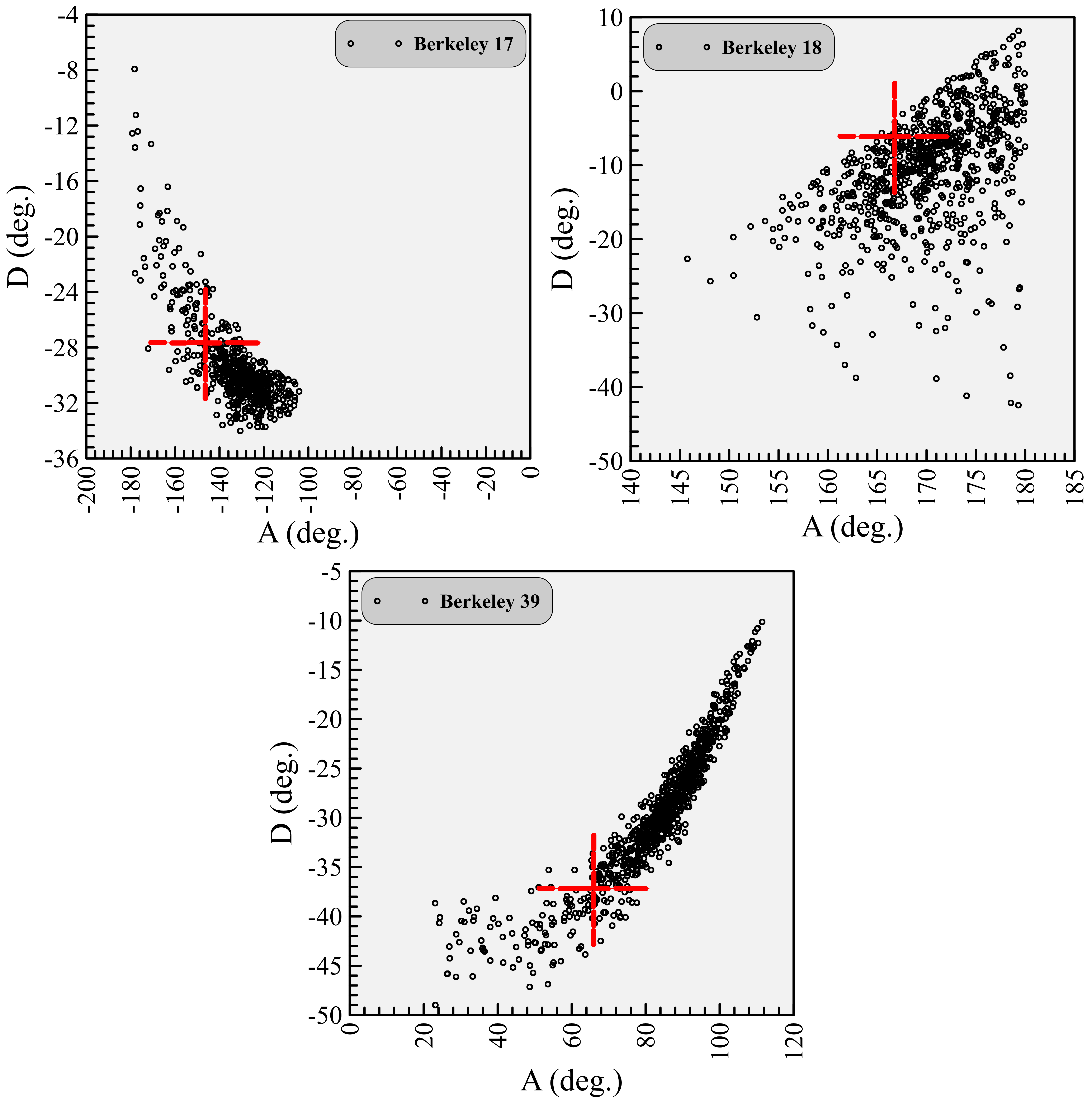}
\caption{AD-diagrams of Be17, Be18, and Be39, where the positions of the computed apex coordinates $(A_0,~D_0)$ are indicated by the cross symbols.}
\label{fig: CP}
\end{figure}

Finding the apex offers crucial information about the cluster's overall dynamical coherence and kinematic structure. This is often accomplished using the AD-diagram approach, which was first presented by \citet{Chupina2001} and \citet{Chupina2006}. With this approach, Figure \ref{fig: CP} displays the apex coordinates ($A_o$, $D_o$) that are derived from the velocity vectors of individual stars. $A_o$ and $D_o$ reflect the convergent point's $\alpha$ and $\delta$, respectively:
\begin{equation}
\label{Eq: A_o and D_o}
A_o = \tan^{-1}\left(\dfrac{\overline{V}_y}{\overline{V}_x}\right),
\end{equation}
\begin{equation}
\label{Eq: 8}
D_o = \tan^{-1} \left(\dfrac{\overline{V}_z}{\sqrt{\overline{V}^{2}_{x} + \overline{V}^{2}_{y}}}\right),
\end{equation}
where the space velocity components along the corresponding axes are denoted by $\overline{V}_x$, $\overline{V}_y$, and $\overline{V}_z$.  Next, the space velocity components $V_x$, $V_y$, and $V_z$ are calculated using the formula provided by \citet{Melchior1958}:
\begin{equation}
\label{Eq. 9-10-11}
\begin{pmatrix}
V_x \\\\
V_y \\\\
V_z
\end{pmatrix} 
= 
\begin{pmatrix}
-4.74~d~\mu_\alpha\cos{\delta}\sin{\alpha} - 4.74~d~\mu_\delta\sin{\delta}\cos{\alpha} \\
+ V_{\rm r} \cos\delta \cos\alpha \\
+4.74~d~\mu_\alpha\cos{\delta}\sin{\alpha} - 4.74~d~\mu_\delta\sin{\delta}\cos{\alpha} \\
+ V_{\rm r} \cos\delta \cos\alpha \\
+4.74~d~\mu_\delta\cos\delta + V_{\rm r}\sin\delta
\end{pmatrix}
\end{equation}

Detailed kinematic and orbital dynamical investigations of open clusters require reliable radial-velocity ($V_R$) measurements. In the present study, we determine the mean radial velocities directly from the stars identified
as members in our astrometric analysis that also have measured radial velocities in the \textit{Gaia}~DR3 archive. The resulting cluster radial velocities are $\langle V_R \rangle = -74.11 \pm 0.27$~km~s$^{-1}$ ($N=22$) for
Be17, $\langle V_R \rangle =-2.30 \pm 0.58$~km~s$^{-1}$ ($N=38$) for Be18, and $\langle V_R \rangle =59.30 \pm 0.16$~km~s$^{-1}$ ($N=30$) for
Be39. For comparison, we got values of the radial velocities from \citet{Hunt2024} which are equal to $-$73.40 $\pm$ 0.86 \textcolor{black}{($N=23$)}, $-$5.50 $\pm$ 2.38 \textcolor{black}{($N=34$)}, and 59.88 $\pm$ 0.63 \textcolor{black}{($N=30$)} km~s$^{-1}$ for Be17, Be18, and Be39, respectively. Our estimates are in good agreement with those of \citet{Hunt2024}.

The space velocity components were translated from the equatorial coordinate system to the Galactic frame using the transformation equations specified in Eqs. (\ref{Eq. 12}). These equations can be used to convert the velocity components ($V_x$, $V_y$, and $V_z$) into their Galactic counterparts, $U$, $V$, and $W$, which stand for the velocity vector components with respect to the Galactic centre. The following formulas were used to determine these components:

\begin{equation}
\label{Eq. 12}
\begin{pmatrix}
U \\\\
V \\\\
W
\end{pmatrix} 
= 
\begin{pmatrix}
-0.0518807421 \; V_{x} -
0.872222642 \; V_{y} -\\
0.4863497200 \; V_{z} \\
+0.4846922369 \; V_{x} -
0.4477920852 \; V_{y} +\\
0.7513692061 \; V_{z}\\
-0.873144899 \; V_{x} -
0.196748341 \; V_{y} +\\
0.4459913295 \; V_{z} \\
\end{pmatrix}
\end{equation}

The mean Galactic space velocity components $\overline{U}$, $\overline{V}$, and $\overline{W}$ are thus calculated by averaging the individual velocity components of each star in the cluster. These average velocities are expressed using the following formulas:

\begin{equation} 
\label{eq: 15}
\overline{U} = \dfrac{1}{N}\sum ^{N}_{i=1}U_i,\;\;
\overline{V} = \dfrac{1}{N}\sum ^{N}_{i=1}V_i, \;\; \text{and} \; \; \overline{W} = \dfrac{1}{N}\sum ^{N}_{i=1}W_i.
\end{equation}

where $N$ represents the total number of stars in the cluster. Furthermore, the average space velocity components $(U,~V,~W)$ for the clusters are shown in Table \ref{tab:full_parameters}. 

\subsubsection{Solar Motion Elements}

Solar motion is the velocity of the Sun relative to the Local Standard of Rest (LSR), a reference frame that depicts the standard motion of neighbouring stars. OCs, composed of stars with a common origin, age, and velocity, are good tracers for determining solar motion because of their uniform kinematics and widespread distribution over the Galactic disk. The solar space velocity components, expressed in km s$^{-1}$, can be computed from the mean spatial velocity components ($\overline{U},~\overline{V},~\overline{W}$) of a certain stellar cluster in Galactic coordinates. These components are determined using the following relations \citep{Elsanhoury2016, Elsanhoury2022, Haroon2025}:

\begin{equation}
\label{Eq.16}
U_{\odot} = -\overline{U}, \; V_{\odot} = -\overline{V}, \; \text{and} \; W_{\odot} = -\overline{W}.
\end{equation}

The magnitude of the solar space velocity ($S_{\odot}$) with respect to the observed objects is then determined using the following expression. On the other hand, the solar apex ($l_\text{A}$, $b_\text{A}$) in Galactic coordinates takes the formulas:

\begin{equation}
\label{Eq. 17}
S_{\odot}=\sqrt{(\overline{U})^2+(\overline{V})^2+(\overline{W})^2}.
\end{equation}
and
\begin{equation}
\label{Eq. 18}
l_\text{A} = \tan^{-1}\left(\frac{-\overline{V}}{\overline{U}}\right) \;\; \text{and} \;\;
b_\text{A} = \sin^{-1} \left(\frac{-\overline{W}}{S_\odot}\right),
\end{equation}

\begin{table*}[ht]
\centering
\caption{The evolving, kinematical, and dynamical parameters.}
\begin{tabular}{ll|ccc}
\toprule
&\textbf{Parameter} & \textbf{Berkeley 17} & \textbf{Berkeley 18} & \textbf{Berkeley 39}  \\
\hline
\hline
\multicolumn{5}{c}{\textbf{Evolving Parameters}} \\
\midrule
&$R_h$ (pc) & 3.46 $\pm$ 0.34 & 7.51 $\pm$ 0.47 & 4.07$\pm$ 0.29  \\
&$T_{\rm relax}$ (Myr)&62$\pm$2 & 199$\pm$7 & 86$\pm$2 \\
&$\tau$ & 146 & 17 & 59 \\
&$\tau_{\rm ev}$ (Myr) & 6200 & 19900 & 8600  \\
\hline
\hline
\multicolumn{5}{c}{\textbf{Kinematical Parameters}} \\
\midrule

&$\overline{V_x}$ (km s$^{-1}$) & -72.11 $\pm$ 8.49 & -78.84 $\pm$ 8.88 & 34.06 $\pm$ 5.84     \\
&$\overline{V_y}$ (km s$^{-1}$) & -47.58 $\pm$ 6.90 & 18.66 $\pm$ 4.32 & 77.77 $\pm$ 8.82     \\
&$\overline{V_z}$ (km s$^{-1}$) & -45.45 $\pm$ 6.74 & -8.65 $\pm$ 2.94 & -64.89 $\pm$ 8.06     \\
&$\overline{U}$ (km s$^{-1}$) & 67.34 $\pm$ 8.21 & -7.98 $\pm$ 2.82 & -38.04 $\pm$ 5.83     \\
&$\overline{V}$ (km s$^{-1}$) & -47.80 $\pm$ 6.91 & -53.07 $\pm$ 7.28 & -67.07 $\pm$ 8.19     \\
&$\overline{W}$ (km s$^{-1}$) & 52.06 $\pm$ 7.22 & 61.31 $\pm$ 7.83 & -73.98 $\pm$ 8.60     \\
&$S_{\rm \odot}$ (km s$^{-1}$) & 97.62 $\pm$ 9.88 & 81.48 $\pm$ 9.03 & 106.86 $\pm$ 10.34     \\
&$U_{\rm LSR}$ (km s$^{-1}$) & 74.57 $\pm$ 0.64 & -33.45 $\pm$ 2.55 & -115.05 $\pm$ 0.27  \\
&$V_{\rm LSR}$ (km s$^{-1}$) & -7.86 $\pm$ 2.18 & 13.79 $\pm$ 0.92 & -48.80 $\pm$ 1.44  \\
&$W_{\rm LSR}$ (km s$^{-1}$) & 36.92 $\pm$ 2.50 & 22.39 $\pm$ 2.24 & -31.20 $\pm$ 4.09  \\
&$S_{\rm LSR}$ (km s$^{-1}$) & 83.58 $\pm$ 3.38 & 42.58 $\pm$ 3.51 & 128.81 $\pm$ 4.34  \\
\hline
\hline
\multicolumn{5}{c}{\textbf{Dynamical Parameters}} \\
\midrule
&$A_o~(^\circ)$ & -146.59 $\pm$ 0.13 & 166.68 $\pm$ 0.17 & 66.35 $\pm$ 0.09  \\
&$D_o~(^\circ)$ & -27.75 $\pm$ 0.18 & -6.10 $\pm$ 0.20 & -37.39 $\pm$ 0.01  \\
&$l_A~(^\circ)$ & 35.36 &  -81.45 & -60.44  \\
&$b_A~(^\circ)$ & -32.23 & -48.80 & 43.82  \\
\hline

&$Z_{\rm max}$ (kpc) & $1.397\pm0.188$ & $1.235\pm0.221$ & $1.421\pm0.257$  \\
&$R_{\rm a}$ (kpc)   & $14.474\pm0.254$ & $16.154\pm0.896$ & $11.855\pm0.403$  \\
&$R_{\rm p}$ (kpc)   & $8.248\pm0.168$  & $13.410\pm0.583$ & $8.318\pm0.177$   \\
&$R_{\rm m}$ (kpc)  & $11.361\pm0.210$ & $14.782\pm0.739$ & $10.086\pm0.289$  \\
&$e$                & $0.274\pm0.001$  & $0.093\pm0.006$  & $0.175\pm0.006$   \\
&$T_{\rm p}$ (Myr)   & $335\pm7$        & $443\pm25$       & $289\pm8$      \\

\bottomrule
\end{tabular}
\label{tab:full_parameters}
\end{table*}

\section{Orbit Analyses}

The derivation of the space-velocity vectors and orbital parameters for the three Berkeley clusters was accomplished using the \textsc{galpy} Python package \citep{Bovy15}, employing the axisymmetric Galactic potential model \textsc{MWPotential2014}. The necessary inputs for this computation included the mean equatorial coordinates ($\alpha, \delta$), mean proper motions ($\langle\mu_{\alpha}\cos\delta, \mu_{\rm \delta}\rangle$), distance ($d_{\rm (m-M)}$), and the newly computed mean radial velocities ($\langle V_{\rm R}\rangle$), along with their respective uncertainties. The adopted solar parameters for the orbital integrations were a Galactocentric distance of $R_{\rm gc} = 8$ kpc and a local circular velocity of $V_{\rm rot} = 220$ km s$^{-1}$ \citep{Bovy15, Bovy12}.

To reconstruct the clusters' dynamical pasts, their trajectories were integrated backward in time for a duration of 3 Gyr with a 1.5 Myr time step, ensuring that multiple complete orbital periods were covered \citep{Cinar2025}. This process yielded key dynamical parameters, including the apogalactic ($R_{\rm a}$) and perigalactic ($R_{\rm p}$) distances, orbital eccentricity ($e$), maximum vertical displacement from the Galactic plane ($Z_{\rm max}$), and the orbital period ($T_{\rm P}$). The heliocentric space-velocity components $(U,V,W)$ of
the clusters were computed using the algorithms and transformation matrices of \citet{Johnson87}, adopting a right-handed coordinate system in which $U$ is directed toward the Galactic centre, $V$ in the direction of Galactic rotation, and $W$ toward the North Galactic Pole. As the clusters are located
well beyond the immediate solar neighbourhood (at distances of $\sim 2.9$–$5.8$ kpc), we applied first-order corrections for the Galactic
differential rotation following the prescription of \citet{Mihalas81}. In this scheme, the in-plane velocity components are corrected by $\Delta U$ and
$\Delta V$ terms that depend on Galactic longitude, latitude, and distance, while the $W$ component remains unaffected. For our three clusters the
resulting corrections are $(dU,~dV) = (+4.57,~-6.91)$ km s$^{-1}$ for Be17, $(+42.09,~-13.48)$ km s$^{-1}$ for Be18, and
$(+84.22,~+7.55)$ km s$^{-1}$ for Be39, indicating that the $U$ component is more strongly affected by differential rotation \textcolor{black}{than $V$\citep{Canbay2025a}}. After applying these corrections to the heliocentric $U$ and $V$ components, the Galactic space velocity vectors were transformed
to the Local Standard of Rest (LSR) by adding the solar space velocity $(U,~V,~W)_\odot$ = (8.83 $\pm$ 0.24, 14.19 $\pm$ 0.34, 6.57 $\pm$ 0.21) km s$^{-1}$
\citep{Coskunoglu11}. The differential rotation corrections and the final LSR corrected components $(U,~V,~W)_{\rm LSR}$ for each cluster are listed in
Table~\ref{tab:full_parameters}.

\begin{figure*}[ht]
    \centering
\includegraphics[width=0.99\linewidth]{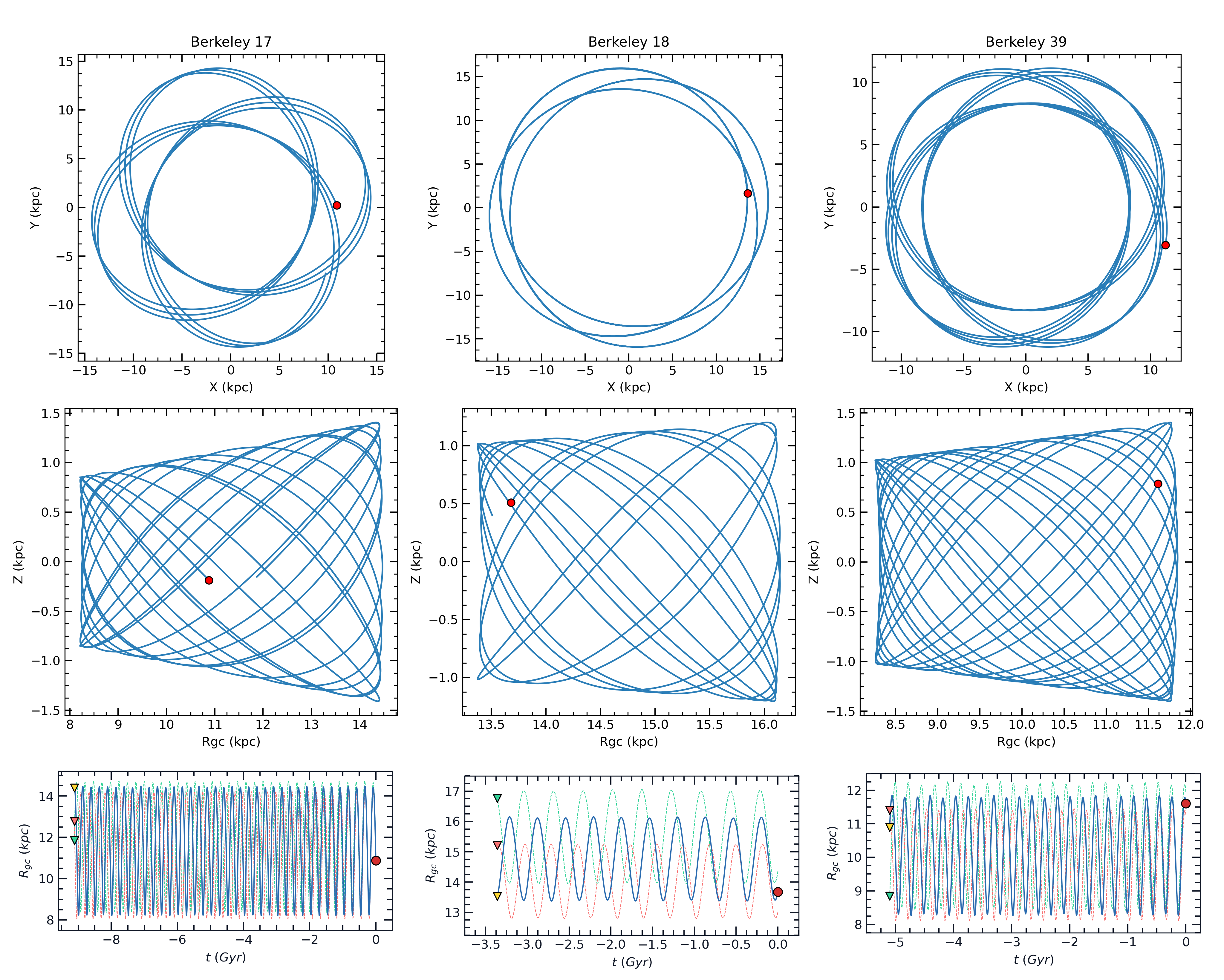}
\caption{Backward integrated orbits for the three open clusters Berkeley 17, 18, and 39. Top panels: Show the projection of the orbits onto the Galactic X-Y plane. Middle panels: Display the motion in the meridional plane ($R_{gc}$ vs. Z). Bottom panels: Illustrate the time evolution of the Galactocentric distance ($R_{gc}$). The red dots in all panels mark the present-day positions of the clusters.}    
\label{Orbits}
\end{figure*}

The resulting orbits, visualized in the $R_{\rm gc}\times Z$ and $R_{\rm gc}\times t$ planes (Figure~\ref{Orbits}), reveal diverse dynamical histories for the sample clusters \citep[e.g.,][]{Tasdemir2025, Karagoz2025, Yucel2024}. The clusters exhibit a wide range of orbital eccentricities. For instance, Be18 has a nearly circular orbit ($e=$ 0.093 $\pm$ 0.006), whereas Be17 and Be39 follow more eccentric paths.

This diversity is crucial for classifying their Galactic population membership. Based on their orbital parameters, a clear distinction emerges. Be17, Be18, and Be39 attain significant heights above the Galactic plane, with $Z_{\rm max}$ values of 1.397 $\pm$ 0.188 kpc, 1.235 $\pm$ 0.221 kpc, and 1.421 $\pm$ 0.257 kpc, respectively. Such large vertical motions are characteristic of the Galactic thick-disk population. Following the classification scheme of \citet{Schuster12}, which primarily uses the $V_{\rm LSR}$ velocity, objects with such significant vertical motions are typically associated with the thick disk. The orbital properties of these three clusters therefore strongly suggest that they are members of the thick disk.

\section{Summary and Conclusions}

In this study, we have performed a comprehensive multifaceted analysis of three intermediate to old age open clusters, Berkeley 17, 18, and 39, leveraging the exceptional precision of the \textit{Gaia} DR3 archive. Our primary goal was to derive a self-consistent and updated set of their fundamental astrometric, photometric, structural, and kinematic parameters, thereby refining their role as tracers of the Milky Way's disk evolution.

A robust membership analysis, founded on the probabilistic methodology of \cite{Yad13}, was applied to distinguish bona fide cluster members from field star contamination. This procedure yielded substantial and statistically significant samples of 600, 1042, and 907 members for Berkeley 17, 18, and 39, respectively. The resulting clean samples were crucial for the precise determination of the clusters' mean properties. The astrometric distances, derived directly from the mean parallaxes, were found to be 3373 $\pm$ 68 pc, 5767 $\pm$ 119 pc, and 4214 $\pm$ 57 pc, respectively. These values are generally in good agreement with recent determinations in the literature, such as those by \cite{Hunt2024} and \cite{Pog21}, but offer improved precision due to our rigorous membership selection.

The structural morphology of each cluster was investigated through radial density profiles (RDPs), which were excellently modeled by the empirical profiles of \cite{King1962}. This analysis provided key structural parameters, including core and limiting radii. Notably, the derived concentration parameters ($C$) indicate a diversity of structural morphologies among the studied clusters, ranging from the moderately concentrated Be18 ($C = 2.84$) to the more centrally condensed Be39 ($C = 3.92$). Be17, with an intermediate concentration ($C = 3.54$), occupies a transitional regime between these two systems. This spread in concentration values suggests that the clusters may have experienced different dynamical evolutionary pathways, potentially reflecting variations in their initial conditions, internal relaxation processes, or the influence of external tidal fields.

By fitting PARSEC isochrones \citep{marigo2017new} to the decontaminated color-magnitude diagrams, we constrained the clusters' ages and reddening. We determined ages of 9.12 $\pm$ 1.00 Gyr (Be17), 3.36 $\pm$ 0.50 Gyr (Be18),and 5.10 $\pm$ 0.50 Gyr (Be39). Our age for Be17 confirms its status as one of the oldest open clusters in the Galaxy, a crucial object for probing the early history of the Galactic disk. For Be39, our age estimate aligns well with values from \cite{Cant20b} but is younger than the age reported by \cite{sampedro2017multimembership}. The reddening values span a wide range, from a modest $E(B-V) = 0.17$ mag for Be39 higher values for Be17 and Be18, reflecting their different lines of sight through the Galactic plane.

The mass functions (MFs) for the MS members were derived, and their slopes ($\alpha$) were calculated. The resulting slopes of 2.44 $\pm$ 0.06 (Be17), 2.72 $\pm$ 0.06 (Be18), and 2.25 $\pm$ 0.07 (Be39) are, within uncertainties, largely consistent with the canonical value of \cite{Salpeter_1955} ($\alpha=2.35$). This consistency suggests that the stellar initial mass function in these clusters was not significantly different from the standard IMF, although some evidence of dynamical evolution and mass segregation may be present. Furthermore, our analysis of the dynamical relaxation times ($T_{\rm relax}$) shows that for all three clusters, their current age is significantly greater than their relaxation time ($\tau = \text{age} / T_{\rm relax} \gg 1$). This confirms that they are dynamically evolved systems where two-body relaxation has had ample time to shape their internal structure.

Combining membership-based mean radial velocities with the derived astrometric parameters allows us to investigate the kinematic and dynamical
properties of the clusters in a self-consistent way. The resulting orbital parameters reveal a clear kinematic differentiation within the sample.
Be17 and Be18 are located in the outer Galactic disc and reach large apogalactic distances of $R_{\rm a} =$ 14.474 $\pm$ 0.254 kpc and
16.154 $\pm$ 0.896 kpc, respectively. Both clusters rise significantly above the Galactic plane; however, Be18 follows an almost circular orbit
($e =$ 0.093 $\pm$ 0.006), whereas Be17 moves on a substantially more eccentric trajectory ($e =$ 0.274 $\pm$ 0.001). Taken together, these results provide a coherent dynamical picture for the three
clusters, refining their fundamental parameters and mapping their distinct orbital paths within the Milky Way.

\section*{Acknowledgements}
We sincerely thank the anonymous referee for their thoughtful comments and helpful suggestions, which greatly enhanced both the clarity and overall quality of this manuscript. This study presents results derived from the European Space Agency (ESA) space mission $Gaia$. The data from $Gaia$ are processed by the $Gaia$ Data Processing and Analysis Consortium (DPAC). Financial support for DPAC is provided by national institutions, primarily those participating in the $Gaia$ Multi-Lateral Agreement (MLA). For additional information, the official $Gaia$ mission website can be accessed at \url{https://www.cosmos.esa.int/gaia}, and the $Gaia$ archive is available at \url{https://archives.esac.esa.int/gaia}. The authors would like to express their gratitude to the Deanship of Scientific Research at Northern Border University, Arar, KSA, for funding this research under project number "NBU-FFR-2026-237-01".

\bibliographystyle{elsarticle-harv}

\end{document}